\documentclass[aps,pra,reprint,superscriptaddress,notitlepage,showpacs,floatfix,twocolumn,longbibliography]{revtex4-1}

\usepackage{graphicx,graphics,epsfig,times,bm,bbm,amssymb,amsmath,amsfonts,mathrsfs}
\usepackage[scr=boondoxo,scrscaled=1.05]{mathalfa}
\usepackage[matrix,frame,arrow]{xy}
\usepackage{hyperref}
\hypersetup{
    colorlinks=true,       % false: boxed links; true: colored links
    linkcolor=red,          % color of internal links
    citecolor=magenta,        % color of links to bibliography
    filecolor=magenta,      % color of file links
    urlcolor=cyan,           % color of external links
    runcolor=cyan
}
\usepackage[normalem]{ulem}
\usepackage[pdftex]{color}
\usepackage{dsfont}
\usepackage[export]{adjustbox}
\usepackage[table]{xcolor}
\usepackage{comment}
\usepackage{algorithm}
\usepackage{algpseudocode}
\usepackage{listings}
\usepackage[caption=false]{subfig}
\newcommand\numberthis{\addtocounter{equation}{1}\tag{\theequation}}

\usepackage{mathtools}

\newcommand{\ket}[1]{ | #1 \rangle  }
\newcommand{\bra}[1]{ \langle  #1 | }
\newcommand{\expect}[1]{\langle #1 \rangle}

\newcommand{\eqq}[1]{Eq.~(\ref{#1})}

\newcommand{\dd}{\mathrm{d}}

\algnewcommand\algorithmicforeach{\textbf{for each}}
\algdef{S}[FOR]{ForEach}[1]{\algorithmicforeach\ #1\ \algorithmicdo}

\begin{document}

\title{Quantum Control Noise Spectroscopy with Optimal Suppression of Dephasing}
\author{Vivian Maloney}\thanks{Vivian.Maloney@jhuapl.edu}
\affiliation{The Johns Hopkins University Applied Physics Laboratory, Laurel, Maryland, 20723, USA}
\author{Yasuo Oda}
\affiliation{William H. Miller III Department of
Physics $\&$ Astronomy, Johns Hopkins University, Baltimore, Maryland 21218, USA}
\author{Gregory Quiroz}
\affiliation{The Johns Hopkins University Applied Physics Laboratory, Laurel, Maryland, 20723, USA}
\affiliation{William H. Miller III Department of
Physics $\&$ Astronomy, Johns Hopkins University, Baltimore, Maryland 21218, USA}
\author{B. David Clader}\thanks{Current affiliation: Goldman Sachs \& Co.}
\affiliation{The Johns Hopkins University Applied Physics Laboratory, Laurel, Maryland, 20723, USA}
\author{Leigh M. Norris}
\affiliation{The Johns Hopkins University Applied Physics Laboratory, Laurel, Maryland, 20723, USA}
\date{\today}

\begin{abstract}
We extend quantum noise spectroscopy (QNS) of amplitude control noise to settings where dephasing noise or detuning errors make significant contributions to qubit dynamics. 
Previous approaches to characterize amplitude noise are limited by their vulnerability to low-frequency dephasing noise and static detuning errors, which can overwhelm the target control noise signal and introduce bias into estimates of the amplitude noise spectrum. To overcome this problem, we leverage optimal control to identify a family of amplitude control waveforms that optimally suppress low-frequency dephasing noise and detuning errors, while maintaining the spectral concentration in the amplitude filter essential for spectral estimation. The waveforms found via numerical optimization have surprisingly simple analytic forms, consisting of oscillating sine waves obeying particular amplitude and frequency constraints. 
In numerically simulated QNS experiments, these waveforms demonstrate superior robustness, enabling accurate estimation of the amplitude noise spectrum in regimes where existing approaches are biased by low-frequency dephasing noise and detuning errors.
\end{abstract}
\maketitle

\section{Introduction}
Robust, accurate control is a requirement for quantum technologies ranging from quantum sensors to quantum computers. In addition to being limited by environmental noise sources or by crosstalk between neighboring quantum systems, current gate fidelities are constrained by noise in the drives and classical electronics used for control \cite{VanDijk2019}. Common examples of control noise include jitter in the amplitude and phase of microwave fields used to implement gates in platforms such as superconducting qubits, trapped ions, quantum dots and diamond NV centers \cite{Soare2014,ball2015walsh,BallMasterClock,Raftery2017}. In numerous platforms, control noise is addressed using composite pulses (CPs), which originated in nuclear magnetic resonance to correct unknown sources of static amplitude and phase noise \cite{Levitt1986}. Approaches such as dynamical decoupling (DD) or dynamically corrected quantum gates (DCGs) are capable of refocusing both control and environmental noise sources, as long as the temporal correlations of the noise decay on timescales that are slow compared to the control \cite{Viola1998,Viola1999,Khodjasteh2009}. While CPs, DD and DCGs are powerful in that they can increase gate fidelities without detailed knowledge of the control noise, further gains can be realized by taking into account specific features of the noise affecting a quantum device and leveraging optimal control to design robust, customized gates \cite{ball2015walsh,Teerawat2021, oda2022fgrafs, 2022TroutOptimalControl}. 

Obtaining the knowledge of the noise necessary for customized error mitigation is the domain of quantum noise spectroscopy (QNS) \cite{Szankowski2017}. Over the last decade, QNS has become a framework to characterize temporally correlated noise in quantum devices by reconstructing the associated noise power spectral densities or noise spectra. The noise spectra provide a means to perturbatively model the noisy dynamics of a quantum device, estimate gate fidelities and define objective functions for optimal control \cite{green2013arbitrary,ball2015walsh}. QNS protocols were originally devised for classical, Gaussian, dephasing noise \cite{Bylander2011,AlvarezPRL2011}, but have been extended to multi-axis noise \cite{PazMultiaxis2019}, non-Gaussian dephasing noise \cite{NorrisPRL2016,Ramon2019} and noise on multiple qubits \cite{Szankowski2016,PazPRA2017,Uwe2020}. Most QNS protocols follow a procedure in which a quantum system is (1) prepared in a state that is sensitive to a target noise source, (2) allowed to evolve under the influence of noise and control, and finally (3) measured to estimate the expected value of an observable that captures the dynamical effect of the noise. The control applied in step (2) modifies the spectral response of the quantum system to noise. Typically, the control is selected so that the system is sensitive to noise at frequencies within a certain range or ``band", yielding an estimate of the noise spectrum in a localized region of the frequency domain.

The first QNS protocol for characterizing quantum control noise, introduced and experimentally validated on trapped ions in Ref. \cite{frey2017NatComm}, utilized control based on Slepian sequences from classical signal processing to estimate the spectrum of multiplicative amplitude noise acting on a single qubit. Waveforms based on Slepian sequences were later extended to enable
simultaneous estimation of amplitude and dephasing noise spectra in Ref.  \cite{frey2020simultaneous}.
Slepian sequences are notable for their optimal spectral concentration or degree of localization in a target frequency band \cite{dpss}. In both the classical and quantum settings, the spectral concentration of Slepians translates into estimates of the noise spectrum with minimal {\it{leakage bias}} or contamination from noise at frequencies outside the target band \cite{dpss,thomson1982spectrum,norris2018optimally}. In addition to leveraging Slepians and other tools from classical signal processing, Refs. \cite{frey2017NatComm} and \cite{frey2020simultaneous} applied a quantum tomographic procedure to isolate the dynamical contribution of the amplitude noise from dephasing, the secondary noise source in the trapped ion device. Because it relies on a perturbative expansion, the tomographic procedure ensures the estimate of the amplitude noise is unbiased by the presence of dephasing provided that both noise sources are sufficiently weak and, furthermore, that the dynamical contribution of the amplitude noise is on the same order or stronger than that of dephasing. While this condition holds in the trapped ion device  owing to the dominance of amplitude noise, higher order terms in the perturbative expansion can become significant in platforms where dephasing is strong relative to amplitude noise and/or predominantly low frequency, characteristic of superconducting qubits and other solid-state devices. In addition to dephasing, higher order contributions can also become significant in the presence of static detuning errors, which can arise from miscalibration or  drift in the carrier frequency of an external control field.

In this paper, we present a protocol for amplitude noise QNS in regimes where dephasing noise or detuning errors make significant contributions to qubit dynamics. Our protocol relies on a novel, reduced-complexity optimization procedure that takes into account realistic device constraints in order to arrive at control waveforms that reduce the impact of higher order perturbative terms while retaining a high degree of spectral concentration. While the optimized waveforms sacrifice some spectral concentration with respect to the Slepians, they effectively suppress low-frequency dephasing noise and cancel static detuning errors. The optimized waveforms are well approximated by a family of analytic functions, which we term the dephasing-robust waveforms. In numerically simulated QNS experiments, we show that the dephasing robust waveforms 
reduce the magnitudes of higher order terms in the perturbative expansion, allowing for reliable estimation of the amplitude noise spectrum when combined with the tomographic procedure of Refs. \cite{frey2017NatComm} and \cite{frey2020simultaneous}.  

The paper is organized as follows. Section \ref{sec::background} summarizes the perturbative treatment of the qubit dynamics under single-axis control, multiplicative amplitude noise and dephasing originally presented in Ref. \cite{frey2017NatComm}. This section also introduces the noise spectrum and control filter functions (FFs), which capture how applied control modifies the response of the qubit to noise in the frequency domain. In Sec. \ref{sec::DephasingRobust}, we present the tomographic procedure, for the first time including perturbative corrections up to fourth order. We then cast the problem of finding a control waveform that minimizes the dominant higher-order corrections into an optimization, using techniques from linear programming to scalably incorporate physical constraints. Lastly, we introduce the analytic dephasing-robust waveforms, which closely approximate the results of our numerical optimization. Finally, in Sec. \ref{sec::QNS}, we compare the dephasing-robust waveforms to the Slepians by numerically simulating amplitude noise QNS on a qubit subject to low-frequency dephasing noise or detuning error at various strengths. As the strength of the amplitude noise decreases relative to the dephasing or detuning error, we demonstrate that the dephasing robust waveforms enable accurate reconstruction of the amplitude noise spectrum.

\section{Background}\label{sec::background}

\subsection{Time domain: control, noise and dynamics}

We consider a single qubit subject to time-dependent, temporally correlated noise and controlled by modulating the amplitude $\Omega(t)$ and phase $\phi(t)$ of an external driving field. By transforming into a frame rotating with the carrier frequency of the drive at resonance with the qubit energy splitting and making the rotating wave approximation, we can write the control Hamiltonian of the qubit in units of $\hbar=1$ as
\begin{equation}\label{eq::Hc}
    H_c(t) = \frac{\Omega(t)}{2} [\cos \phi(t) \sigma_1 + \sin \phi(t) \sigma_2].
\end{equation}

Here, we denote the Pauli operators by $\vec{\sigma}=(\sigma_1,\sigma_2,\sigma_3)\equiv(\sigma_x,\sigma_y,\sigma_z)$. The actual qubit evolution differs from the ideal dynamics generated by $H_c(t)$ due to the presence of dephasing and amplitude control noise. In the rotating frame, the noise is described by the  Hamiltonian \cite{ball2015walsh}
\begin{equation}\label{eq::NoiseH}
    H_N(t) = \beta_{z}(t) \sigma_3 + \beta_{\Omega}(t) H_c(t).
\end{equation}
Here, the dephasing $\beta_{z}(t)$ and amplitude noise $\beta_{\Omega}(t)$ are independent, stationary, Gaussian stochastic processes.
Note that the dephasing enters $H_N(t)$ as an additive frequency fluctuation whereas the amplitude noise acts multiplicatively on the control Hamiltonian, inducing fluctuations in the Rabi frequency with magnitude $\Omega(t)\beta_{\Omega}(t)$. While $\beta_{z}(t)$ has units of frequency, $\beta_{\Omega}(t)$ is dimensionless. 
Static detuning errors, which generate coherent rotations about $\sigma_3$, contribute to the mean of the dephasing process $\beta_z(t)$. 
For later convenience, we write the noise Hamiltonian in matrix form
\begin{align}\label{eq::HNQuadForm}
H_N(t) =\vec{B}(t)\,\mathbf{N}(t)\,\vec{\sigma}^\intercal,
\end{align}
where
\begin{align}
\vec{B}(t)
\label{eq::BVec}
\equiv\,[\beta_{\Omega}(t),\beta_{\Omega}(t),\beta_z(t)]\notag
\end{align}
and
\begin{align}
\mathbf{N}(t)\equiv \begin{pmatrix}
        \frac{1}{2} \Omega(t) \cos\phi(t) & 0 & 0 \\
        0 & \frac{1}{2} \Omega(t) \sin\phi(t) & 0 \\
        0& 0 & 1
    \end{pmatrix}
\end{align}
are the noise vector and noise matrix, respectively.

The complete qubit dynamics, including both control and noise, is described by the rotating-frame Hamiltonian $H(t)=H_c(t)+H_N(t)$. To isolate the dynamical contribution of the noise, we make one additional transformation into the toggling frame or interaction picture associated with $H_c(t)$. If $U_c(t)=\mathcal{T}_+\text{exp}[-i\int_0^tds\,H_c(s)]$ is the ideal gate implemented by the noiseless control, the qubit dynamics in the toggling frame are generated by the ``error Hamiltonian"
\begin{align}
    \tilde{H}(t) = U_c(t)^{\dagger} H_N(t) U_c(t). 
\end{align}
Since it is a unitary conjugation of \eqq{eq::HNQuadForm}, the error Hamiltonian can be expressed in matrix form as
\begin{align}
    \tilde{H}(t) = \vec{B}(t)\,\mathbf{Y}(t)\,\vec{\sigma}^\intercal
    \label{effective-hamiltonian}, 
\end{align}
where the control matrix is defined as $\mathbf{Y}(t)\equiv\mathbf{N}(t)\mathbf{R}(t)$ with $[\mathbf{R}(t)]_{ij}\equiv \frac{1}{2} \textrm{Tr} [ U_c(t)^{\dagger} \sigma_i U_c(t) \sigma_j ] $ \cite{norris2018optimally}. The unitary generated by the error Hamiltonian or ``error propagator", $\tilde{U}(t) = \mathcal{T}_{+} \exp [ -i \int_{0}^{t} \mathrm{d}s \ \tilde{H}(s)]$, evolves the qubit in the toggling frame and is related to the rotating-frame propagator by $U(t)\equiv \mathcal{T}_{+} \exp [ -i \int_{0}^{t} \mathrm{d}s\,H(s)]=U_c(t)\tilde{U}(t)$. Observe that the rotating and toggling frames are equivalent at a time $t$ such that $U_c(t)=I$. 
Following Ref. \cite{green2013arbitrary}, we parametrize the error propagator as
\begin{equation}\label{eq::a}
\tilde{U}(t) = \exp[-i \vec{a}(t) \cdot \vec{\sigma}]=e^{-i\sum_{m}\vec{a}^{(m)}(t) \cdot \vec{\sigma}}, 
\end{equation}
where  $\vec{a}(t) \equiv [a_1(t), a_2(t), a_3(t)]$ is the real, time-dependent ``error vector". On the right-hand side of this expression, the $m^{\text{th}}$ term in the summation  depends on the $m^{\text{th}}$-order term of a perturbative Magnus expansion of the propagator \cite{Magnus}. For a toggling-frame Hamiltonian along multiple Pauli components, it is generally impossible to determine an exact expression for $\vec{a}(t)$. For sufficiently weak noise \cite{green2013arbitrary}, the error vector is approximated by the leading order term in the Magnus expansion, so that $\tilde{U}(t) \approx \exp[-i\int_0^t ds \tilde{H}(s)]$ and
\begin{align}\label{eq::aQuadForm}
\vec{a}(t)\approx \vec{a}^{(1)}(t)\equiv
\int_0^t ds\,\vec{B}(s)\,\mathbf{Y}(s),
\end{align}
where the second equality follows from \eqq{effective-hamiltonian}. 

Thus far, we have considered the most general form of the control Hamiltonian in \eqq{eq::Hc}. In the remainder of this work, we restrict the drive phase to $\phi(t)\in\{0,\pi\}$ so that the control generates rotations about the $x$-axis and 
\begin{equation}\label{eq::HcAmplitude}
H_c(t)=\frac{\Omega(t)}{2}\sigma_1,
\end{equation}
where we have redefined the ``amplitude" control waveform $\Omega(t)$ to take both positive and negative values. For this simplified control,
\begin{align}
\mathbf{N}(t)=&\,\begin{pmatrix}
        \frac{1}{2} \Omega(t) & 0 & 0 \\
        0 & 0 & 0 \\
        0& 0 & 1
    \end{pmatrix},\notag\\\notag\\
\mathbf{R}(t) =&\, \begin{pmatrix}
        1 & 0 & 0 \\
        0 & \cos \Theta(t) & - \sin \Theta(t) \\
        0 & \sin \Theta(t) & \cos \Theta(t)
    \end{pmatrix},\notag\\\notag\\
\mathbf{Y}(t) =&\,  
    \begin{pmatrix}
       \frac{1}{2}\Omega(t) & 0 & 0 \\
        0 & 0 & 0 \\
        0 & \sin \Theta(t) & \cos \Theta(t)
    \end{pmatrix}
   ,\label{eq::YMatrixComponents}
\end{align}
where $\Theta(t) = \int_{0}^{t} \mathrm{d}s \ \Omega (s)$.
From Eqs. (\ref{eq::BVec}), (\ref{eq::aQuadForm}) and (\ref{eq::YMatrixComponents}), the components of the error vector to leading order in the Magnus expansion are \cite{frey2017NatComm,norris2018optimally}
\begin{align}\label{eq::a1}
    a_1^{(1)}(t) &=\frac{1}{2}\int_{0}^{t} \!\!\mathrm{d}s \,  \Omega (s) \beta_{\Omega}(s), \\
    a_2^{(1)}(t) &=\int_{0}^{t}\!\! \mathrm{d}s \, \sin \Theta (s) \beta_{z}(s), \\
    a_3^{(1)}(t) &=\int_{0}^{t}\!\! \mathrm{d}s \ \cos \Theta (s) \beta_{z}(s).\label{eq::a3}
\end{align}
Due to the sparsity of the control matrix, each component of the error vector is the integrated product of a single control matrix element and a single noise source. Observe that the amplitude noise enters the qubit dynamics solely through the $a_1^{(1)}(t)$ component.

\subsection{Frequency Domain: Spectra and Filter Functions}
Because we are ultimately interested in the spectral properties of the noise, we transform from the time domain to the frequency domain, where the noise and control are represented by spectra and filter functions, respectively.
If the dephasing and amplitude noise sources are wide-sense stationary, their autocovariances can be parametrized in terms of the \emph{lag time}, $\tau\equiv t_2-t_1$, so that
\begin{align*}
\left \langle\Delta \beta_n (t_1) \Delta\beta_n (t_2) \right \rangle = \left \langle\Delta \beta_n (\tau)\Delta \beta_n (0) \right \rangle
\end{align*}
where $n\in\{\Omega,z\}$, $\Delta O\equiv O-\expect{O}$, and $\left \langle\cdot\right\rangle$ denotes the ensemble average over realizations of the amplitude and dephasing noise processes. By the Weiner-Kinchin theorem, the power spectral density (PSD) or ``spectrum" of the noise is the Fourier transform of the noise autocovariance with respect to the lag time,
\begin{align}
    S_{n}(\omega) = \frac{1}{2\pi} \int_{-\infty}^{\infty}\!\! \mathrm{d}\tau\  \left \langle  \Delta\beta_n (\tau) \Delta\beta_n (0) \right \rangle e^{-i \omega \tau}. \label{PSD}
\end{align}
The objective of our noise spectroscopy procedure is estimating $S_\Omega(\omega)$, the spectrum associated with the amplitude noise.

Estimating $S_\Omega(\omega)$ requires experimentally measurable quantities that depend on the noise spectra. We take the approach of Ref. \cite{frey2017NatComm} in which the amplitude noise spectrum is inferred from quantities derived from the error vector components. For notational convenience, define
\begin{align}\label{eq::beta_vec}
\vec{\beta}(t)&\equiv [\beta_{\Omega}(t),\beta_z(t),\beta_z(t)],\\
\vec{y}(t)&\equiv
[\Omega(t)/2,\sin \Theta(t),\cos \Theta(t)],
\end{align}
so that the error vector components in Eqs. (\ref{eq::a1})-(\ref{eq::a3}) are given by $a_i^{(1)}(t) =\int_{0}^{t} \mathrm{d}s\, y_i(s) \beta_i(s)$ for $i\in\{1,2,3\}$. The variances of the error vector components are then \cite{frey2017NatComm,norris2018optimally}
\begin{align}
    \langle \Delta a_i^{(1)}(T)^2 \rangle &= \int_{0}^{T} \!\!\!\mathrm{d}t_2 \!\int_{0}^{T}\!\!\! \mathrm{d}t_1\, \big\langle \Delta\beta_i(t_1) \Delta\beta_i(t_2)  \big\rangle\,  y_i(t_1) y_i(t_2)\notag\\
    &=\frac{1}{\pi}\int_{0}^\infty\!\!d\omega\,S_i(\omega)\,F_i(\omega,T).\label{eq::overlap}
 \end{align}   
This expression is an overlap integral between the noise spectrum,
\begin{align*}
S_i(\omega)\equiv\begin{cases} S_\Omega(\omega),&i=1,\\
S_z(\omega),&i=2,3,
\end{cases}
\end{align*}  
and the associated \emph{control filter function} (FF) \cite{KofmanKurizkiFF,CywinskiDD,ball2015walsh},
\begin{align}
F_i(\omega,T)\equiv\left|\int_0^T\!\!\!dt\,e^{i\omega t}y_i(t)\right|^2.
\end{align}
The control FFs are frequency-domain representations of the control generated by $H_c(t)$. For later convenience, we follow Ref. \cite{ball2015walsh} and group the FFs according to whether they describe the qubit's response to amplitude or dephasing noise in \eqq{eq::overlap}, forming the amplitude and dephasing FFs 
\begin{align}
    F_{ \Omega }( \omega , T) &\equiv F_{ 1 }( \omega , T) =
    \frac{1}{4}\left|\int_0^T\!\!\!dt\,e^{i\omega t}\,\Omega(t)\right|^2, \label{eq::AmpFF}\\\notag\\
    F_{Z}( \omega , T) &\equiv F_{ 2 }( \omega , T) + F_{ 3 }( \omega , T)\notag\\&=\!
    \left|\int_0^T\!\!\!\!\!dt\,e^{i\omega t}\sin\Theta(t)\right|^2\!\!\!+\left|\int_0^T\!\!\!\!\!dt\,e^{i\omega t}\cos\Theta(t)\right|^2\!\!.\label{eq::ZFF}
\end{align}
\noindent
Observe that the amplitude noise spectrum enters the qubit dynamics through $\langle \Delta a_1^{(1)}(T)^2 \rangle$, while the dephasing noise spectrum is present in both $\langle \Delta a_2^{(1)}(T)^2 \rangle$ and $\langle \Delta a_3^{(1)}(T)^2 \rangle$, hence the two contributions to the dephasing FF in \eqq{eq::ZFF}.

\subsection{Quantum Noise Spectroscopy}
QNS protocols characterize the environmental noise of a quantum system by measuring system observables under various operational settings and correlating the measured values to the strength of different noise sources. The filter function formalism (FFF) offers a particularly clear insight into how this process works. In the FFF, to leading order in a perturbative expansion, the average value of a measured system quantity is expressed as an overlap integral between the spectrum of a noise source and a control FF,
\begin{align}
\expect{M(T)}=C\int_{0}^\infty\!\!d\omega\, S(\omega) F(\omega,T),
\end{align}
where $C$ is some proportionality constant.

Note that the variance in \eqq{eq::overlap} takes precisely this form. Applying different controls to the quantum system modifies the shape of $F(\omega,T)$ in the frequency domain. For QNS, the FF is often engineered to be localized within a chosen passband or frequency interval, e.g. $B\equiv[\omega_0-\delta\omega,\omega_0+\delta\omega]$. For sufficiently small $\delta\omega$, $S(\omega)$ is approximately constant within $B$. With the additional assumption that the support of $F(\omega,T)$ is minimal outside $B$, the quantity $\expect{M(T)}$ is approximately proportional to $S(\omega_0)$,
\begin{align}\label{eq::OverlapLinear}
\expect{M(T)}\approx S(\omega_0)\,C\int_{B}\!\!d\omega\, F(\omega,T).
\end{align}
Since $C$ and $F(\omega,T)$ are known, we can infer $S(\omega_0)$.
Choosing a set of control waveforms that concentrate $F(\omega,T)$ in different passbands then enables us to characterize $S(\omega)$ over a range of frequencies. 

Note that this procedure depends critically on $F(\omega,T)$ being {spectrally concentrated} or localized in the target frequency band, $B$. If $F(\omega,T)$ exhibits {spectral leakage} or out-of-band spectral components, the linear approximation of the overlap integral in \eqq{eq::OverlapLinear} is no longer valid and $\expect{M(t)}$ will depend on the value of the spectrum outside of $B$. Refs. \cite{frey2017NatComm,norris2018optimally,frey2020simultaneous} minimized spectral leakage for QNS applications by using control waveforms derived from Slepian sequences, a family of discrete sequences that maximize spectral concentration for spectral estimation based on classical time series \cite{thomson1982spectrum}. In absence of suitable control to localize the FF, the spectrum can still be estimated through the frequency comb techniques \cite{AlvarezPRL2011}, linear inversion or equivalent approaches that discretize the overlap integral over a range of frequencies \cite{FerrieNJP2018,murphy2021universal}. Spectral leakage is still a detriment in these settings, however, as delocalized filter functions
can produce non-sparse, potentially ill-conditioned linear systems that amplify error in the spectral estimate. As such, maximizing spectral concentration and minimizing spectral leakage is an important principle of filter design for QNS.

\section{Dephasing-robust characterization of amplitude noise}\label{sec::DephasingRobust}
Since $\langle\Delta a_1^{(1)}(T)^2 \rangle$ is an overlap integral between the amplitude noise spectrum and the amplitude filter function, the filter-shaping strategies of the previous section can, in principle, be applied to infer $S_\Omega(\omega)$. Accomplishing this in a realistic system, however, requires a means of experimentally measuring $\langle\Delta a_1^{(1)}(T)^2 \rangle$. For noise that is zero-mean, Ref.  \cite{frey2017NatComm} devised a strategy to measure $\langle\Delta a_1^{(1)}(T)^2 \rangle$ that holds in a weak noise limit. As the strength of the noise increases, however, our ability to reliably determine
$\langle\Delta a_1^{(1)}(T)^2 \rangle$ is complicated by terms that are higher order in the noise strength. In particular, higher order dephasing dependent terms can make it impossible to isolate amplitude noise dependent quantities, biasing estimates of $S_\Omega(\omega)$.

\subsection{Bias from dephasing noise}
For weak, zero-mean noise and amplitude control that generates a net identity (ensuring the toggling and rotating frames are equivalent), Ref. \cite{frey2017NatComm} determined 
$\langle\Delta a_1^{(1)}(T)^2 \rangle$ from a tomographic procedure depending on three experimentally measurable survival probabilities,
\begin{align}\label{eq::axMeas}
\langle\Delta a_1^{(1)}(T)^2 \rangle &\approx \frac{1}{2} \left [ 1 \!+ \mathbb{P}(\uparrow_1, T) \! - \mathbb{P}(\uparrow_2, T) \! - \mathbb{P}(\uparrow_3, T) \right ]
\equiv  \mathcal{P}. \end{align}
  Here, $\mathbb{P}(\uparrow_i, T) = \langle  \lvert \bra{\uparrow_i\!} {U}(T) \ket{\!\uparrow_i} \rvert ^2 \rangle$ is the  probability that the qubit remains in state $\ket{\!\uparrow_i}$ after evolving under noise and control for a time $T$ and $\ket{\!\uparrow_i}$ denotes the +1 eigenstate of $\sigma_i$. This estimate holds in a weak-noise limit in which dynamical contributions beyond second order in $\beta_z(t)$ and $\Omega(t)\beta_\Omega(t)$ are negligible. 
  Beyond this limit, terms that are higher order in the error vector components are no longer negligible in \eqq{eq::axMeas}. In Appendix \ref{app::Measurements}, we extend the measurement strategy to noise with nonzero mean and include perturbative terms up to fourth order in $\beta_z(t)$ and $\Omega(t)\beta_\Omega(t)$. In the main text, we focus on dephasing noise with nonzero mean in order to treat static detuning errors, but take the amplitude noise to be zero-mean for simplicity. With higher-order perturbative terms and nonzero-mean dephasing, \eqq{eq::axMeas} becomes
  \begin{align}\notag
  &\langle \Delta a_1^{(1)}(T)^2 \rangle-\frac{1}{3}\langle a_1^{(1)}(T)^4 \rangle\,+\langle a_1^{(2)}(T)^2 \rangle\notag
  \\&-\frac{1}{3}\langle a_2^{(1)}(T)^2\,a_1^{(1)}(T)^2\rangle
  -\frac{1}{3}\langle a_3^{(1)}(T)^2\,a_1^{(1)}(T)^2\rangle\notag\\
  &\;\;\;\;\;\;\;\;\;\approx \frac{1}{2} \left [ 1 + \mathbb{P}(\uparrow_1, T) - \mathbb{P}(\uparrow_2, T) - \mathbb{P}(\uparrow_3, T) \right ]. \label{eq::axMeasHigherOrder}
  \end{align}
Here, the superscript $(2)$ denotes the second order Magnus contribution to the error vector in \eqq{eq::a}, an explicit expression for which is provided in Appendix \ref{app::Measurements}.
The first two terms of \eqq{eq::axMeasHigherOrder} depend solely on amplitude noise, while the remaining terms depend either on dephasing noise or both amplitude and dephasing noise. When the dephasing noise is strong relative to the amplitude noise or when the control increases the sensitivity of the qubit to dephasing noise or detuning error, the terms on the second line can dominate over the other higher order terms and become comparable in magnitude to $\langle\Delta a_1^{(1)}(T)^2 \rangle$. Therefore, na\"ively employing \eqq{eq::axMeas} for amplitude noise spectroscopy will result in an estimate of $S_\Omega(\omega)$ that is biased by dephasing.

To better elucidate the relationship between noise strength, noise spectra, amplitude control, and the relative magnitudes of the terms in \eqq{eq::axMeasHigherOrder}, we analyze the higher order terms in the frequency domain. First, define the overlap integrals
\begin{align}\label{eq::In}
&I_\Omega(T)\equiv\,\frac{1}{2\pi}\int_{-\infty}^\infty \!\!\!\! d\omega\,F_\Omega(\omega,T)S_\Omega(\omega),\\
&I_Z(T)\equiv\,\frac{1}{2\pi}\int_{-\infty}^\infty \!\!\!\! d\omega\,F_Z(\omega,T)\big[S_z(\omega)+2\pi\mu_z^2\delta(\omega)\big],\label{eq::Iz}
\end{align} 
where $\langle\beta_z(t)\rangle\equiv\mu_z$ is static in time due to stationarity.
Observe that $\big\langle \Delta a_1^{(1)}(T)^2 \big\rangle =I_\Omega(T)$.
Since $\beta_\Omega(t)$ and $\beta_z (t)$ are Gaussian, their fourth-order moments factor into products of one- and two-point correlation functions. Using this factorization, we show in Appendix \ref{app::Measurements}, that the higher order terms take the form
\begin{align}
&\big\langle a_1^{(1)}(T)^4 \big\rangle =3\;I_\Omega(T)^2,\label{eq::a14}\\
&\big\langle a_1^{(1)}(T)^2a_2^{(1)}(T)^2\big\rangle
+\big\langle a_1^{(1)}(T)^2a_3^{(1)}(T)^2\big\rangle=I_\Omega(T)I_Z(T),\label{eq::IomIz}\\\notag
&\big\langle a_1^{(2)}(T)^2 \big\rangle=\!\int_{-\infty}^\infty\!\!\!\!d\omega\!\!\int_{-\infty}^\infty\!\!\!\!d\omega' \,G_Z(\omega,\omega',T)\Big[\,\frac{S_z(\omega)\,S_z(\omega')}{(2\pi)^2}\\&+\frac{\mu_z^2}{2\pi}S_z(\omega)\delta(\omega')+\frac{\mu_z^2}{2\pi}S_z(\omega')\delta(\omega)
+\frac{\mu_z^4}{3}\delta(\omega)\delta(\omega')\Big]\label{eq::a22}.
\end{align}
Here, $G_Z(\omega,\omega',T)$ is a higher order FF given by
\begin{align}\label{eq:GZ}
&G_Z(\omega,\omega',T)=\int_0^{T}\!\!\!\!dt_1\!\!\int_0^{t_1}\!\!\!\!dt_2\,\sin\big[\Theta(t_1)\!-\!\Theta(t_2)\big]\\
&\;\;\;\times\int_0^{T}\!\!\!\!dt_3\!\!\int_0^{t_3}\!\!\!\!dt_4\,\sin\big[\Theta(t_3)\!-\!\Theta(t_4)\big]\notag\big[e^{i\omega(t_1-t_2)}e^{i\omega'(t_3-t_4)}\\&\;\;\;\;\;\;\;\;\;+e^{i\omega(t_1-t_3)}e^{i\omega'(t_2-t_4)}
+e^{i\omega(t_1-t_4)}e^{i\omega'(t_2-t_3)}\big].\notag
\end{align}
While \eqq{eq::a14} depends quadratically on the overlap integral between the amplitude noise spectrum and the amplitude FF, \eqq{eq::IomIz} is the product of the overlap integrals associated with amplitude noise and dephasing noise. Dephasing noise also enters \eqq{eq::a22}, which is a two-dimensional overlap integral involving the dephasing noise spectrum and a higher order FF.
We can express \eqq{eq::axMeasHigherOrder} in terms of the overlap intergrals as 
\begin{align}
&\frac{1}{2} \left [ 1 + \mathbb{P}(\uparrow_1, T) - \mathbb{P}(\uparrow_2, T) - \mathbb{P}(\uparrow_3, T) \right ] \label{eq:axMeasHigherOrderCompact}\\
&\approx I_\Omega(T) - I_\Omega(T)^2 - \frac{1}{3} I_\Omega(T) I_Z(T) + \big\langle a_1^{(2)}(T)^2 \big\rangle \notag,
\end{align}
which highlights the source of each higher-order correction. If \eqq{eq::axMeas} is used to estimate $\big\langle \Delta a_1^{(1)}(T)^2 \big\rangle =I_\Omega(T)$, then \eqq{eq::IomIz} contributes a dephasing-dependent \textit{multiplicative} bias proportional to $I_Z(T)$. The magnitude of this bias can be estimated from the measured survival probabilities,
\begin{align}
I_Z(T) &= \langle \Delta a_2^{(1)}(T)^2 \rangle + \langle \Delta a_3^{(1)}(T)^2 \rangle \notag
\\
&\approx
  1 - \mathbb{P}(\uparrow_1, T) . \label{eq::ayzMeasr}
\end{align}
The remaining higher-order term in \eqq{eq:GZ} contributes a dephasing-dependent \textit{additive} bias.

The additive and multiplicative biases  can contribute to the estimator when the dephasing noise is stronger than the amplitude noise on the interval $[0,T]$, i.e.
\begin{align}
\int_0^T\!\!dt\,\big|\beta_z(t)\big|^2>\int_0^T\!\!dt\,\big|\Omega(t)\beta_\Omega(t)\big|^2,
\end{align}
or when the dephasing FFs $F_Z(\omega,T)$ and $G_Z(\omega,\omega',T)$ have significant spectral support in regions of the frequency domain where $S_z(\omega)$ and $S_z(\omega)S_z(\omega')$ are respectively large. In this later case, the magnitudes of $I_Z(T)$ and $|\expect{a_1^{(2)}(T)^2}|^{1/2}$ can be larger than $I_\Omega(T)$ even when the dephasing noise strength is comparable to or weaker than the amplitude noise strength. A particularly relevant example involves the $1/f$ dephasing noise present in solid state platforms such as superconducting qubits and semiconductor qubits. Since $\mu_z\neq0$ in the presence of detuning error, detuning contributes additional bias through \eqq{eq::Iz} when the dephasing FF has DC support, i.e. $F_Z(0,T)\neq 0$. Similarly, if $G_Z(\omega,\omega',T)\neq 0$ at $\omega=0$ or $\omega'=0$, detuning error can also contribute bias through \eqq{eq::a22}.

\subsection{Optimized Waveforms}
When the FFs have low-frequency or DC support, the higher-order dephasing dependent terms can make a significant contribution to \eqq{eq::axMeasHigherOrder}. As a consequence, $\langle\Delta a_1^{(1)}(T)^2 \rangle$ can no longer be isolated using the measurement strategy of Ref.  \cite{frey2017NatComm}, biasing estimates of the amplitude noise spectrum. 
Extending amplitude noise QNS to settings with detuning error or  dephasing noise that is predominantly low frequency requires a new control strategy. To engineer amplitude control waveforms that suppress low frequency dephasing noise and detuning error while maintaining spectral concentration of the amplitude filter, we turn to optimal control. 
The problem of filter design with bounded controls can be cast as a scalable optimization problem. Our objective is to design control waveforms of the form in \eqq{eq::HcAmplitude}  suited for amplitude noise spectroscopy in the presence of low-frequency dephasing noise or detuning error. As such, the amplitude FF should be spectrally concentrated and the dephasing FF should act as a high-pass filter, capable of ``filtering out" low-frequency and DC spectral components. To accomplish this, we exploit the properties of the family of Slepian or discrete prolate spheroidal sequences (DPSS)  \cite{dpss}. For completeness, we have provided a brief summary of the DPSS and their properties in Appendix \ref{app:DPSS}.  The DPSS of order $k$ is denoted by $\{v_n^{(k)}(N,W)\,|\,n\!=\!0,\ldots,N\!-\!1\}$, where $N$ is the length of the sequence and $W\in(0,1/2)$ is the bandwidth parameter determining the size of the passband in the frequency domain. Owing to the spectral concentration of their discrete-time Fourier transforms, the lowest-order DPSS have been used extensively in classical spectral estimation.  
Refs. \cite{frey2017NatComm, norris2018optimally} harnessed the spectral concentration of the DPSS for amplitude noise spectroscopy in the quantum setting by employing the piecewise constant amplitude waveforms
\begin{align}
\Omega(t)=\begin{cases}
\Omega\,v_0^{(k)}(N,W),& t\in[0,\Delta t),\\
\Omega\,v_1^{(k)}(N,W),& t\in[\Delta t,2\Delta t),\\
\;\;\;\;\;\;\;\;\;\;\;\;\;\;\;\;\vdots&\;\;\;\;\;\;\;\;\;\;\vdots\\
\Omega\,v_{N-1}^{(k)}(N,W),& t\in[(N-1)\Delta t,N\Delta t),
\end{cases}\label{eq::DPSSwaveform}
\end{align} 
where $\Omega$ is the control amplitude in units of angular frequency and $\Delta t$ is a time increment.
This waveform produces an amplitude FF that is spectrally concentrated in the passband $[-2\pi W/\Delta t,2\pi W/\Delta t]$. Through analog modulation techniques, the passband can be shifted to different locations along the frequency axis \cite{ norris2018optimally}. For example, sine modulation entails modifying the waveform above so that $\Omega(t)=\Omega\sin(\omega_0 n\Delta t)v_n^{(k)}(N,W)$ for $t\in[n\Delta t,(n+1)\Delta t)$. This produces an amplitude FF with a positive-frequency passband $B_\text{DPSS}(\omega_0)\equiv[-2\pi W/\Delta t+\omega_0,2\pi W/\Delta t+\omega_0]$. Cosine modulation, in which $\sin(\omega_0 n\Delta t)$ is replaced in the amplitude control waveform by $\cos(\omega_0 n\Delta t)$, likewise produces an amplitude FF with positive-frequency passband $B_\text{DPSS}(\omega_0)$.

Inspired by the GRAFS approach of Refs. \cite{GRAFS, oda2022fgrafs}, we parametrize the amplitude control waveform as a piecewise constant linear combination of cosine and sine modulated DPSS,
\begin{widetext}
\begin{align}
\Omega(t)=\begin{cases}
\sum_{k=0}^{K-1}\Omega_\text{c}^{(k)}v_0^{(k)}(N,W),& t\in[0,\Delta t),\\
\sum_{k=0}^{K-1}\Big\{\Omega_\text{c}^{(k)}\cos[\omega_0\Delta t]+\Omega_\text{s}^{(k)}\sin[\omega_0\Delta t]\Big\}\,v_1^{(k)}(N,W),& t\in[\Delta t,2\Delta t),\\
\;\;\;\;\;\;\;\;\;\;\;\;\;\;\;\;\;\;\;\;\;\;\;\;\;\;\;\;\;\;\;\;\;\;\;\;\;\;\;\;\;\;\vdots&\;\;\;\;\;\;\;\;\;\;\vdots\\
\sum_{k=0}^{K-1}\Big\{\Omega_\text{c}^{(k)}\cos[\omega_0(N\!-\!1)\Delta t]+\Omega_\text{s}^{(k)}\sin[\omega_0(N\!-\!1)\Delta t]\Big\}\,v_{N-1}^{(k)}(N,W),& t\in[(N\!-\!1)\Delta t,N\Delta t).
\end{cases} \label{eq:piecewise-constant-waveform}
\end{align} 
\end{widetext}
 Here, the contributions of the cosine and sine modulated DPSS of order $k$ are determined, respectively, by the real amplitudes $\Omega_\text{s}^{(k)}$ and $\Omega_\text{c}^{(k)}$. Since for all $\phi \in \mathbb{R}$ there exist $\Omega_\text{s}^{(k)}$ and $\Omega_\text{c}^{(k)}$ such that $\sin (\omega_0 m \Delta t + \phi) = \Omega_\text{s}^{(k)} \sin (\omega_0 m \Delta t ) + \Omega_\text{c}^{(k)} \cos (\omega_0 m \Delta t )$,
we do not lose any expressive capability by restricting the phase to zero in \eqq{eq:piecewise-constant-waveform}. Due to the linearity of the Fourier transform, this waveform produces an amplitude FF spectrally concentrated in $B_\text{DPSS}(\omega_0)$, equivalent to the passband of the constituent cosine and sine modulated DPSS. 
 
 While spectral concentration of the amplitude filter is enforced by the DPSS, we achieve the desired high-pass dephasing FF by minimizing a control dependent objective function over the $2K$-dimensional space of the amplitudes $\vec{\Omega}\equiv[\Omega_\text{c}^{(0)},\Omega_\text{s}^{(0)},\ldots,\Omega_\text{c}^{(K-1)},\Omega_\text{s}^{(K-1)}]$. Since $\Omega(t)$ is restricted to low order DPSS for which $2K\ll N$, this optimization offers a significant dimensionality reduction over approaches such as GRAPE that optimize over control parameters in each of the $N$ timesteps \cite{GRAFS}. To suppress low-frequency dephasing noise, we choose an objective function given by the overlap integral in \eqq{eq::In} with a $1/f$ dephasing noise spectrum,
\begin{equation}
I_Z(\vec{\Omega},T)=\frac{1}{\pi}\int_{0}^{\infty} \!\!\!d\omega \ F_{Z}( \omega,\vec{\Omega}, T)\, \frac{1}{\omega + \delta \omega} \label{eq:objective},
\end{equation}
where $\delta \omega>0$ is a small regularization term to make the integral well-defined.
Minimizing this objective function corresponds to suppressing the fourth order, dephasing-dependent terms in \eqq{eq::IomIz}, reducing the bias from dephasing noise. While we do not explicitly incorporate the fourth order contribution from \eqq{eq::a22} into the objective function, we found that the waveforms resulting from our optimization reduce the magnitude of this term, as we will demonstrate numerically.

\begin{figure}
    \centering
    \begin{minipage}{0.48\textwidth}
        \centering
        \includegraphics[width=1.0\textwidth]{\detokenize{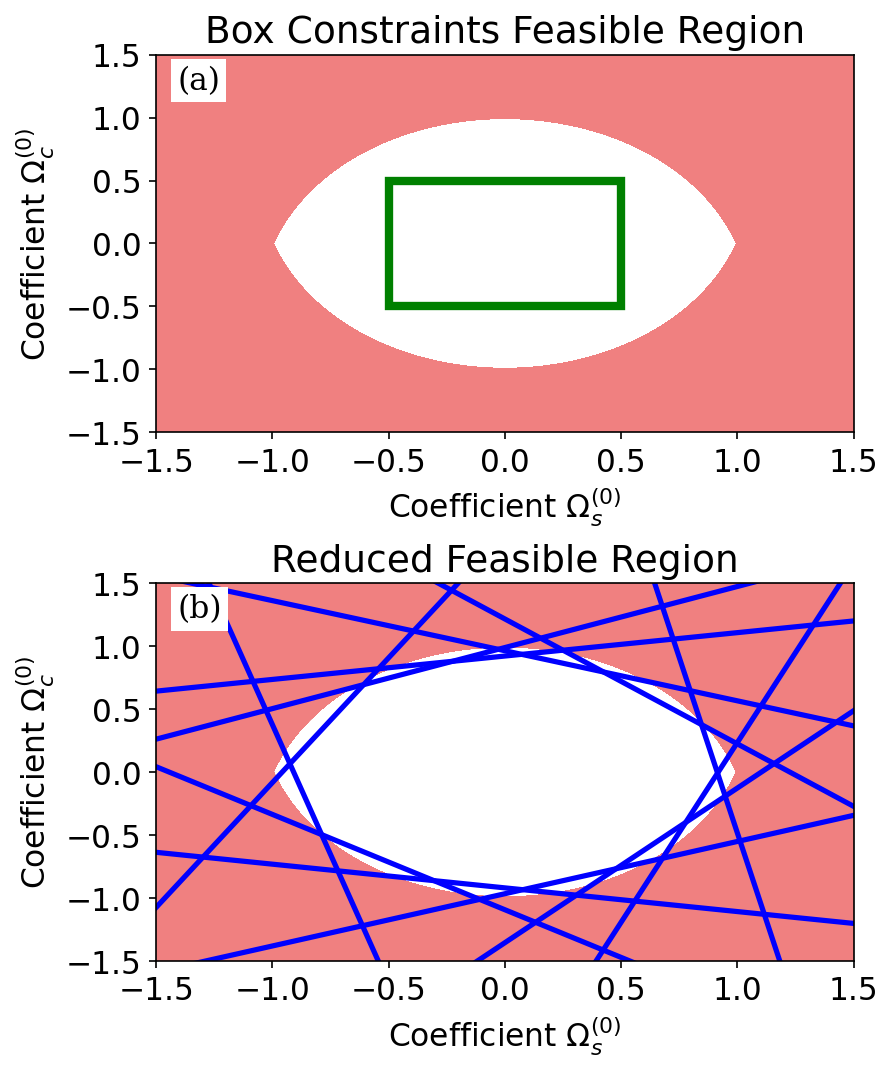}} 
        \caption{Amplitude constraints for optimal control in a toy example with $K=1$ and $\Omega_\textrm{max}=1$. Plot (a) compares the feasible region produced by box constraints (the interior of the green rectangle) to the true feasible region (white), showing that large portions of the true feasible region are excluded by the box constraints. In (b), the linear-programming reduced feasible region (area enclosed by the blue lines) is a much closer approximation of the true feasible region. }
        \label{fig:box-feasible}
    \end{minipage}
\end{figure}
 
In realistic experimental implementations, the control amplitude is bounded by the maximum Rabi rate, $\Omega_\textrm{max}$.
This imposes an $L_{\infty}$ constraint on the control waveform in \eqq{eq:piecewise-constant-waveform},
\begin{align}\notag
\bigg|\sum_{k=0}^{K-1}\!
v_{m}^{(k)}\!(N,W)
\big\{\Omega_\text{c}^{(k)}\!\cos[\omega_0 m \Delta t]
+\Omega_\text{s}^{(k)}&\sin[\omega_0 m \Delta t]\big\}\bigg|\\&\leq  \Omega_\textrm{max}
 \label{eq:a-constraints}
\end{align}
for $m\in\{0,\ldots,N-1\}$. The set of $\vec{\Omega}$ satisfying these constraints defines the true feasible region of physically realizable, piece-wise constant control waveforms.
Directly imposing all $2N$ inequality constraints in our optimization, however, is not scalable and removes the low-dimension advantage of the Slepian functional basis. 
Reference \cite{GRAFS} instead accounts for the maximum Rabi rate by imposing box inequality bounds on the individual basis function coefficients, 
\begin{align*}
  K\max_m \lvert v_{m}^{(k)}(N,W) \rvert \,\lvert\Omega_\text{s}^{(k)}\rvert &\leq \frac{\Omega_\textrm{max}}{2}, 
   \\
  K\max_m \lvert v_{m}^{(k)}(N,W) \rvert \,\lvert\Omega_\text{c}^{(k)}\rvert &\leq \frac{\Omega_\textrm{max}}{2}, 
\end{align*}
 for $k\in\{0,\ldots,K-1\}$ and using the L-BFGS-B optimization algorithm, which natively handles such constraints. This approach reduces the number of constraints from $2N$ to {$4K$}.
However, as depicted in Fig.~\ref{fig:box-feasible}(a) for the 2-dimensional case with $K=1$, the feasible region defined by the box constraints disregards large portions of the true feasible region. 
Note that for this example, the corners of the box constraint do not touch the edges of the true feasible region.
An optimization algorithm will fail to find solutions that lie within the true feasible region if they fall outside the box constraints. This is particularly problematic for high-amplitude waveforms, which lie at the boundary of the true feasible region.

To reduce the number of constraints without overly restricting the feasible region, we developed a dimensionality reduction procedure based on linear programming. This procedure efficiently approximates the true feasible region with a reduced set $I$ of linear inequality constraints numbering $\lvert I \rvert \ll 2 N$. Intuitively, the dimensionality reduction identifies the most critical constraints as those with a significant effect on the size of the feasible region if added or removed. It retains these critical constraints while removing the constraints which have a minor impact on the feasible region. The critical constraints that remain in set $I$ take the form 
\begin{align}
    \forall i\in I\quad  \sum_{k=0}^{K-1} \left ( b_{\text{s}, i}^{(k)} \Omega_{\text{s}}^{(k)} + b_{\text{c}, i}^{(k)} \Omega_{\text{c}}^{(k)} \right ) \leq \Omega_\text{max}
    \label{eq:b-constraints},
\end{align}
where $b_{\text{s}, i}^{(k)}, b_{\text{c}, i}^{(k)}\in\mathbb{R}$.
Further details on the optimization and dimensionality reduction procedure are given in Appendix \ref{app:Optimization}, which may be of independent interest for linear programming optimization problems with a large number of affine constraints. 
 
Crucially, if the waveform in \eqq{eq:piecewise-constant-waveform} satisfies this set of reduced linear constraints, then it also satisfies the set of original constraints in \eqq{eq:a-constraints}. That is, the feasible region generated by the reduced linear constraints is a subset of the true feasible region. A comparison of the feasible regions generated by the box constraints and the reduced linear constraints along with the true feasible region for $K=1$ is shown in Fig.~\ref{fig:box-feasible}. In this example, the $|I|=12$ reduced linear constraints generate a feasible region that closely approximates the true feasible region generated by the $N=40,000$ constraints in \eqq{eq:a-constraints}, a dramatic improvement over the box constraints. This advantage continues into higher dimensions with $K>1$.

\begin{figure*}
\begin{subfloat}{}
    \includegraphics[clip,width=\textwidth]{\detokenize{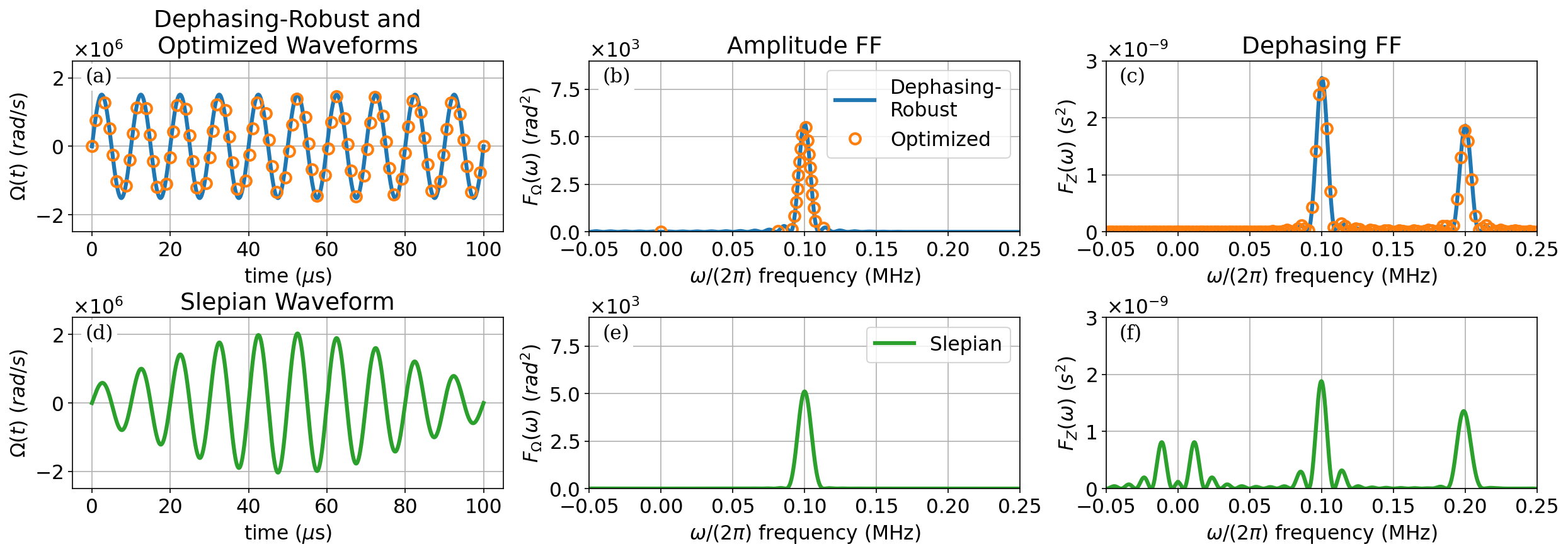}}
   \caption{Comparison of waveforms and FFs for amplitude noise QNS. The optimized waveforms (orange circles) and dephasing-robust waveforms corresponding to the first root of $J_0$ (blue solid line) are plotted (a) with their associated amplitude FFs (b) and dephasing FFs (c). The optimized and dephasing-robust waveforms and FFs almost completely overlap. Shown in (d) is the $k=0$ DPSS (Slepian) waveform for $N=2\times 10^4$ and $NW=1$ along with the corresponding amplitude (e) and dephasing (f) FFs. All waveforms have total time duration $T=100\,\mu$s and are modulated at frequency $\omega_0/2\pi=\lambda/2\pi=0.1$ MHz to shift the passband of the amplitude FF. }
    \label{fig:big-ff-plots}
\end{subfloat}
\end{figure*}

In addition to the waveform amplitude being bounded by the maximum Rabi rate, the control must generate a net identity in order for \eqq{eq::axMeasHigherOrder} to hold. We can ensure that the control implements an identity gate by requiring that the amplitude waveform  integrates to zero, i.e. $\int_0^Tdt\,\Omega(t)=0$. For our piecewise-constant waveforms in \eqq{eq:piecewise-constant-waveform}, this condition is imposed by 
\begin{align}\label{eq::cconstraints}
\sum_{k=0}^{K-1} &\Big[\Omega_\text{s}^{(k)}c_\text{s}^{(k)} + \Omega_\text{c}^{(k)}c_\text{c}^{(k)} \Big]= 0,
\end{align}
where
\begin{align*}
c_\text{s}^{(k)} &= \sum_{m=0}^{N-1} \sin[\omega_0 m \Delta t] v_{m}^{(k)}(N,W),
\\
c_\text{c}^{(k)} &= \sum_{m=0}^{N-1} \cos[\omega_0 m \Delta t] v_{m}^{(k)}(N,W).
\end{align*}
We further add a nonlinear constraint that the DC component of the dephasing FF is exactly zero, $F_Z(\omega=0,\vec{\Omega},T)=0$, which eliminates the contribution of detuning error in \eqq{eq::Iz}.
This constraint can be discretized and expanded as
\begin{align}
    0 &= \left|\sum_{i=0}^{N-1}\sin\Theta_i\right|^2+\left|\sum_{i=0}^{N-1}\cos\Theta_i\right|^2\!\! , \label{eq::Fz00}
\end{align}
 where
\begin{align*}
\Theta_i &= \Delta t \sum_{m=0}^{i-1}\! \sum_{k=0}^{K-1}\!
v_{m}^{(k)}\!(N,W)
\big\{\Omega_\text{c}^{(k)}\!\cos[\omega_0 m \Delta t]
\\&\qquad \qquad \qquad\qquad\qquad\; + \Omega_\text{s}^{(k)} \sin[\omega_0 m \Delta t]\big\}  .
\end{align*}
Equations (\ref{eq::cconstraints}) and (\ref{eq::Fz00}) add two additional constraints to the $|I|$ linear constraints required to bound the Rabi rate. 

We implemented the non-linear optimization using parameters relevant to superconducting transmon qubits. In \eqq{eq:piecewise-constant-waveform}, we took $N=20,000$ and $\Delta t=5$ ns, so that the total duration of the waveform is $T=N\Delta t=100 \ \mu$s. For the DPSS, we selected the bandwidth parameter so that $NW=1$.   
Typical applications use only the first $2 NW$ DPSS, as higher order Slepians are less spectrally concentrated. In our optimization, we found that using the first $2NW + 1$ DPSS, or equivalently taking $K=3$ in \eqq{eq:piecewise-constant-waveform}, produced superior results at the expense of slightly increased spectral leakage in the amplitude FF. Using a gradient-based interior-point solver \cite{IPOPT, casadi, gnu-parallel}, 
we minimized the objective function $I_z(\vec{\Omega},T)$ over $\vec{\Omega}$ subject to the linear constraints in Eqs. (\ref{eq:b-constraints}) and (\ref{eq::cconstraints}).
We performed this optimization for 200 different modulation frequencies $\omega_0\in\{2\pi/T,2\cdot2\pi/T,\ldots,\, 200\cdot2\pi/T\}$, so that $\omega_0/2\pi$ ranged from $0.01$ MHz to $2.00$ MHz. For each frequency $\omega_0$, the modulated DPSS basis functions in \eqq{eq:piecewise-constant-waveform} define a new family of parameterized waveforms, a new feasible region (and reduced feasible region), and a new optimization problem
\begin{equation}
\begin{aligned}
\min_{\Omega_\text{c}^{(k)},\Omega_\text{s}^{(k)}} \quad & 2\int_{0}^{\infty} d\omega \ F_{Z}( \omega,\vec{\Omega}, T) \frac{1}{\omega + \delta \omega}
\\
\textrm{s.t.} \quad &
\forall i \in I\quad  \sum_{k=0}^{K-1} \Big(b_{\text{s}, i}^{(k)} \Omega_{\text{s}}^{(k)} + b_{\text{c}, i}^{(k)} \Omega_{\text{c}}^{(k)} \Big) \leq \Omega_\text{max}
,\\
  & \sum_{k=0}^{K-1} \Big(\Omega_\text{s}^{(k)}c_\text{s}^{(k)} + \Omega_\text{c}^{(k)}c_\text{c}^{(k)} \Big)= 0
,\\
  &F_{Z}( \omega=0,\vec{\Omega} , T) = 0
.
\end{aligned}
\label{eq:NLP}
\end{equation}
A sample waveform resulting from this optimization at modulation frequency $\omega_0/2\pi=0.1$ MHz  is displayed in Fig.~\ref{fig:big-ff-plots} along with the associated amplitude and dephasing FFs. For comparison, Fig.~\ref{fig:big-ff-plots} also shows the FFs generated by the $k=0$ DPSS (Slepian) waveform in \eqq{eq::DPSSwaveform} combined with sine modulation at frequency $\omega_0/2\pi=0.1$ MHz. The amplitude FFs of both the optimized  [Fig.~\ref{fig:big-ff-plots}(b)] and the Slepian [Fig.~\ref{fig:big-ff-plots}(e)] are spectrally concentrated in passbands centered at $0.1$ MHz. Although the optimized amplitude FF displays more spectral leakage in the form of oscillatory ``sidelobes" outside the passband, the optimized dephasing FF in Fig.~\ref{fig:big-ff-plots}(c) is largely suppressed at frequencies below $0.1$ MHz. This is a substantial improvement over the dephasing FF generated by the Slepian waveform, which has significant support at low frequencies. This demonstrates the effectiveness of the optimization in reducing the low-frequency components of the dephasing FF.

\subsection{Analytic Dephasing-Robust Waveforms}
The optimized waveform for $\omega_0/2\pi=0.1$ MHz obtained in the previous section and plotted in Fig.~\ref{fig:big-ff-plots}(a) appears to be a simple oscillatory function with a relatively constant amplitude. This pattern continues over a range of modulation frequencies. Specifically, we find that
the numerically optimized waveforms are well-approximated by analytic functions, which we term ``dephasing-robust" waveforms. These waveforms are defined by
\begin{equation}
    \Omega_\text{d-r}(t) \equiv \Omega_0 \, \sin(\lambda t) \label{magical-time-domain},
\end{equation}
where the amplitude $\Omega_0$ and control modulation frequency $\lambda$ are constrained to take particular values that enforce suppression of low-frequency dephasing noise. Namely, $\Omega_0$ and $\lambda$ must satisfy
\begin{align}
&\lambda=\frac{2\pi M}{T},\label{eq::lambdaCondition}\\
&J_0\!\left(\frac{\Omega_0}{\lambda}\right)=0,\label{eq::J0Condition}
\end{align}
where $M$ is an integer, $T$ is the total duration of the waveform, and $J_0$ is the order-0 Bessel function of the first kind. Interestingly, a sinusoidal driving field satisfying these same conditions was employed in Ref. (\cite{HansenPRA2021}) to create dressed qubits robust to detuning and static amplitude offsets. The first condition guarantees that the dephasing-robust waveforms generate a net identity over the time interval $[0,T]$, as $\int_0^Tdt\,\Omega_\text{d-r}(t)=0$.
The second condition is equivalent to requiring that the ratio $\Omega_0/\lambda$ be a root of $J_0$. The Bessel function $J_0$ has infinitely many roots, with the first three being approximately 2.40, 5.52 and 8.65. As we will show, the first and second conditions imply that $F_Z(0,T)=0$, ensuring that the mean-dependent contributions in Eqs. (\ref{eq::Iz}) and (\ref{eq::a22}) vanish.
Figure~\ref{fig:big-ff-plots}(a) shows that the dephasing-robust waveform corresponding to the first root is nearly identical to the waveform obtained by numerical optimization. 

\begin{figure*}[ht]
\begin{subfloat}{}
    \includegraphics[clip,width=\textwidth]{\detokenize{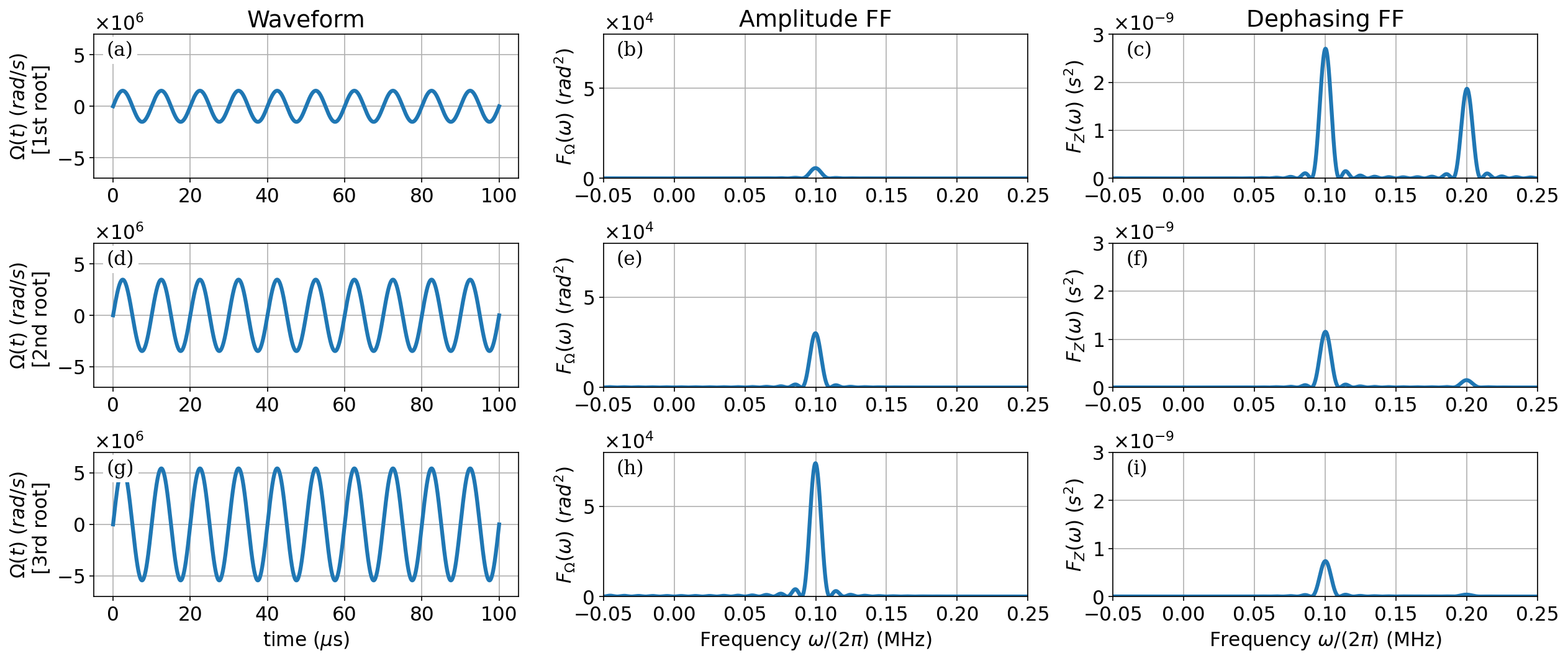}}
    \caption{Leading order dephasing-robust waveforms and FFs.
    Dephasing-robust waveform solutions corresponding to the first (top row), second (middle row) and third (bottom row) roots of $J_0$, all with modulation frequency $\lambda=0.10$ MHz and total time duration $T=100\,\mu$s. Plotted from left to right are the amplitude control waveforms (a,d,g), the amplitude FFs (b,e,h) and dephasing FFs (c,f,i). Note that the dephasing FFs are suppressed at frequencies below $\lambda$.
    }
    \label{fig:3-bessel-ff}
\end{subfloat}
\end{figure*}

As desired for amplitude noise QNS, the amplitude FFs generated by the dephasing-robust waveforms are localized in tunable passbands set by $\lambda$. 
The amplitude FF follows from \eqq{eq::AmpFF} and takes the form
\begin{align}
    F_{\Omega}(\omega,T) &= \left ( \frac{\Omega_0 \lambda \sin(\omega  T/2)}{\omega^2 - \lambda^2 } \right )^2, \label{magical-aff_1}
\end{align}
which is spectrally concentrated in the passband
$B_\text{d-r}(\lambda)=[\lambda-2\pi/T,\lambda+2\pi/T]$. 

The most notable property of the dephasing-robust waveforms is that spectral concentration of the amplitude FF is preserved, while low frequency components of the dephasing FF are suppressed. Using the periodic structure of the dephasing-robust waveform, we show in Appendix \ref{app::DephasingFF} that the dephasing filter can be expressed as 
\begin{align}\notag
F_Z(\omega,T) &= \frac{\sin^2(M\pi\omega/\lambda)}{\sin^2(\pi\omega/\lambda)}\Bigg[\;\bigg|\!\int_0^{2\pi/\lambda}\!\!\!\dd t\cos\Theta(t)e^{-i\omega t}\bigg|^2\\&\;\;\;\;\;\;\;\;\;\;\;\;+\bigg|\!\int_0^{2\pi/\lambda}\!\!\!\dd t\sin\Theta(t)e^{-i\omega t}\bigg|^2\;\Bigg].
\label{magical_deph_ff_1}
\end{align}
The significance of \eqq{eq::J0Condition} becomes apparent when we examine this expression at $\omega=0$, noting that
\begin{align}
&\int_0^{2\pi/\lambda}\!\!\!\dd t\cos\Theta(t)
=\frac{2\pi\cos(\Omega_0/\lambda)J_0(\Omega_0/\lambda)}{\lambda},
\\&\int_0^{2\pi/\lambda}\!\!\!\dd t\sin\Theta(t)=\frac{2\pi\sin(\Omega_0/\lambda)J_0(\Omega_0/\lambda)}{\lambda}.
\end{align}
The condition  $J_0(\Omega_0/\lambda)=0$, thus,  guarantees $F_Z(0,T)= 0$, which cancels detuning error. 

The term immediately to the right of the equality in \eqq{magical_deph_ff_1}, which is known as the Fej\'er kernel \cite{PandW}, is a periodic function with peaks centered at integer multiples of $\lambda$. For fixed $\lambda$, the height of the peaks increases and the width of the peaks decreases with increasing $M$. When $M\gg1$, the Fej\'er kernel is approximated by a frequency comb or sum of delta functions centered at integer multiples of $\lambda$ \cite{AlvarezPRL2011}. In this regime, we show in Appendix \ref{app::DephasingFF} that the dephasing FF takes the form 
\begin{align}
F_Z(\omega,T)
&\approx 2 \pi T \sum_{k\in \mathbb{Z}}\delta(\omega-k\lambda) \left |J_{k}(\tfrac{\Omega_0}{\lambda}) \right|^2, \label{magical-ff}
\end{align}
where $J_k$ is the order-$k$ Bessel function of the first kind. Since $F_Z(0,T)= 0$,
the dephasing FF is a high-pass filter that suppresses low frequency dephasing noise up to the cutoff frequency $\lambda$.

Along with the formal derivation of \eqq{magical-ff} presented in Appendix \ref{app::DephasingFF}, there is an intuitive explanation for the structure of the dephasing FF based on an analogy with dynamical decoupling sequences. 
We note that the dephasing-robust waveforms in \eqq{magical-time-domain} consist of an alternating train of half-sine window pulses, which for the lowest Bessel root correspond to a rotation of approximately $\pm 1.53 \pi$. This is reminiscent of dynamical decoupling sequences with finite-width $\pm \pi$-pulses  \cite{viola2003robust, ishikawa2018influence}.
Using \eqq{eq::ZFF}, the zero-frequency component of the dephasing FF can be expressed as
\begin{align}
    F_Z(\omega=0, T) &=  \left|\int_{0}^{T} \mathrm{d}t \exp i\Theta(t)\right|^2.  \label{eq:dd_FZ_w_zero}
\end{align}
For an alternating train of $\pm \pi$ finite-width pulses, $\Theta(t)$ smoothly transitions between $[0, \pi]$, meaning that $\exp i\Theta(t)$ is always within the first two quadrants of the complex plane. As such, $\exp i\Theta(t)$ has a positive imaginary component for all $t$ and $F_Z(\omega=0, T)$ in \eqq{eq:dd_FZ_w_zero} will be nonzero.
Over-rotating the pulses beyond $\pi$, as is done by the dephasing-robust waveforms, allows $\Theta(t)$ to take values beyond $[0, \pi]$ and  $\exp i\Theta(t)$ to take values in all four quadrants of the complex plane. With a carefully chosen over-rotation, contributions from quadrants III and IV cancel those from quadrants I and II, which causes the dephasing FF to vanish at zero-frequency. With the DC component of the dephasing FF removed, the structure of the waveform as an evenly-spaced alternating pulse train produces a high-pass filter similarly to dynamical decoupling sequences.

\subsection{Comparison of Waveforms}
Since the dephasing-robust and optimized waveforms are nearly identical for the first root of $J_0$, it is unsurprising that the amplitude and dephasing FFs in Fig.~\ref{fig:big-ff-plots}(b) and \ref{fig:big-ff-plots}(c) also closely match. Figure~\ref{fig:3-bessel-ff} shows the dephasing-robust waveforms and FFs for the first three roots of $J_0$ with $\lambda/2\pi=0.1$ MHz. For all roots, the amplitude FFs are spectrally concentrated about $\lambda$. The high-pass nature of the dephasing FF with $\lambda$ as the low-frequency cutoff is also evident for all roots. Interestingly, the high-frequency support of $F_Z(\omega,T)$ beyond $\lambda$ differs between the roots with the second nonzero harmonic at $2\lambda$ being considerably smaller for the second and third roots. Importantly, as evidenced by the low-frequency support of $F_Z(\omega,T)$ for the DPSS waveforms and the absence of low-frequency support of $F_Z(\omega,T)$ for the dephasing-robust waveforms, the three leading-order roots offer superior suppression of low-frequency dephasing noise. 

\begin{figure}[h]
    \centering
    \includegraphics[width=0.4\textwidth]{\detokenize{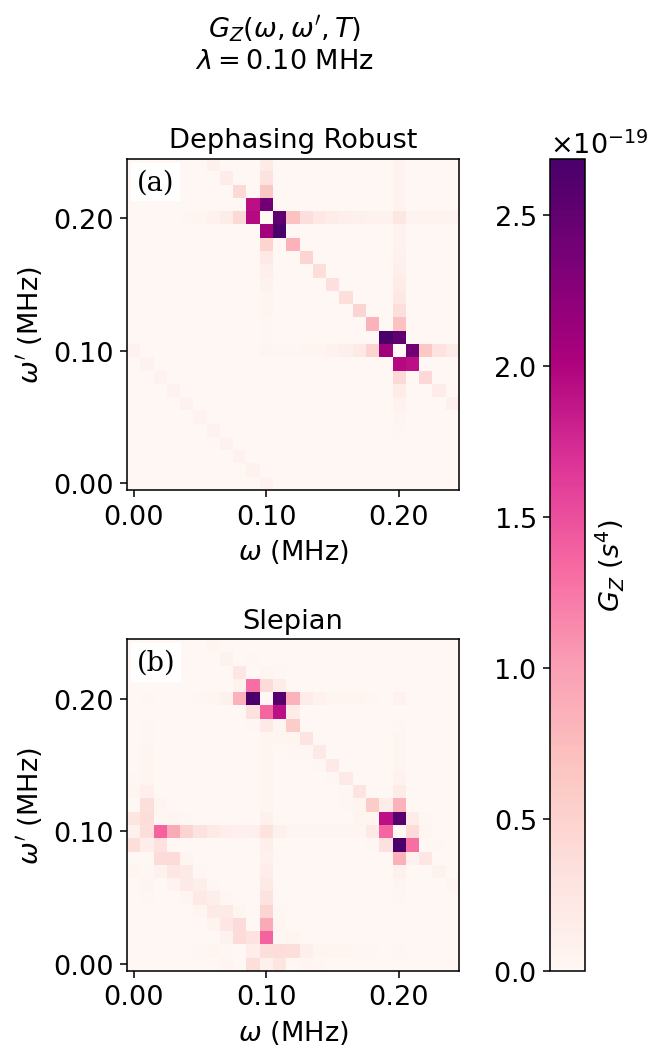}} 
    \vspace*{-10pt}
    \caption{The higher-order dephasing filter function $G_z(\omega, \omega', T)$ for the Dephasing-Robust (a) and Slepian (b) waveforms with $\lambda = 0.10$ MHz and $T = 100 \; \mu s$. The unwanted peak at $(0, \lambda)$ in the bottom-left corner of the Slepian FF demonstrates a sensitivity to a combination of low-frequency and static dephasing, while the absence of this peak for the dephasing-robust indicates that the waveforms are robust against higher-order low-frequency dephasing noise.
    }
    \label{fig:Gz}
\end{figure}

While we have thus far focused on the noise filtering properties of $F_Z(\omega,T)$, the strength of the dephasing-dependent higher order terms in \eqq{eq::axMeasHigherOrder} also depends on the higher order FF $G_Z(\omega,\omega',T)$ defined in \eqq{eq:GZ}. Unlike $F_Z(\omega,T)$,  $G_Z(\omega,\omega',T)$ forms a two-dimensional surface in the frequency domain. 
The higher-order perturbative term in \eqq{eq::a22} illustrates that low-frequency dephasing noise is suppressed when $G_Z(\omega,\omega',T)$ has minimal low-frequency support over both $\omega$ and $\omega'$. Similarly, detuning error is suppressed when $G_Z(\omega,\omega',T)=0$ at $\omega=0$ and $\omega'=0$.
Figure~\ref{fig:Gz}(a) and \ref{fig:Gz}(b), respectively, show   $G_Z(\omega,\omega',T)$ generated by the dephasing-robust waveform in Fig.~\ref{fig:big-ff-plots}(a) and the DPSS waveform in Fig.~\ref{fig:big-ff-plots}(d). To produce these plots, we evaluated the quadruple integral that defines $G_Z(\omega,\omega',T)$ using an efficient iterated fast Fourier transform procedure described in Appendix \ref{app::DFT}.
For the dephasing-robust waveform,
$G_Z(\omega,\omega',T)$ is dominated by two peaks at $(\omega,\omega')=(\lambda,2\lambda)$ and $(\omega,\omega')=(2\lambda,\lambda)$ and has
minimal spectral support when $\omega,\,\omega'<\lambda$.  Similar to $F_Z(\omega,T)$, the modulation frequency $\lambda$ acts as a cutoff below which low-frequency dephasing noise is suppressed. The DPSS waveform, on the other hand, has more support at low and DC frequency with additional peaks at $(\omega,\omega')=(0,\lambda)$ and $(\omega,\omega')=(\lambda,0)$, making it vulnerable to detuning error when the dephasing noise spectrum has support at $\omega\approx\lambda$. The high-pass nature of $G_Z(\omega,\omega',T)$ for the dephasing-robust waveforms again translates into improved filtering of low-frequency dephasing noise and detuning error.

\section{Application to Amplitude Noise QNS}\label{sec::QNS}
The dephasing-robust waveforms closely match the results of our numerical optimization and, therefore, offer near-optimal suppression of low-frequency dephasing and detuning error while at the same time preserving spectral concentration in the amplitude FF. The high-pass filtering characteristics of the FFs $F_Z(\omega,T)$ and $G_Z(\omega,\omega',T)$ ensure that the dephasing-robust waveforms minimize the dynamical contributions of both low-frequency dephasing and detuning error. The resulting reduction in dephasing-induced bias ensures that the tomographic measurement strategy can reliably isolate the dynamical contribution of amplitude noise. 
To quantify the practical difference this makes for noise characterization, we numerically simulate amplitude noise QNS in the presence of low-frequency dephasing noise and detuning errors using both the  dephasing-robust and DPSS waveforms.

To assess the robustness of the waveforms in the presence of low-frequency dephasing versus detuning, we implement two different numerical QNS experiments in which the qubit evolves under amplitude noise and either (1) static detuning error or (2) low frequency dephasing. 
To fairly compare the performance of the waveforms for a range of modulation frequencies, we consider a flat amplitude noise spectrum with a high-frequency cutoff $\omega_h$, 
\begin{align*}
 &S_\Omega(\omega)=\begin{cases}
A_\Omega, &  \omega \leq \omega_h, \\
      0, & \omega > \omega_h.
\end{cases}
\end{align*}
Static detuning error is introduced by taking $\beta_z(t)=\Delta$ in the noise  
Hamiltonian of \eqq{eq::NoiseH}. For the dephasing, we simulate noise with a $1/f$ spectrum, characteristic of solid state devices \cite{Paladino2014}, with both low and high-frequency cutoffs,

\begin{align*}
S_z(\omega)=
C  
\begin{cases}
\frac{A_z}{\omega_l}, &   \omega\leq \omega_l, \\
      \frac{A_z}{\omega}, & \omega_l < \omega \leq \omega_h,\\
      0, & \omega > \omega_h.
\end{cases}
\end{align*}
Here, the low-frequency cutoff $\omega_l$ ensures that the strength of the dephasing noise is finite at $\omega=0$ and the dimensionless scale factor $C$ determines the noise strength. 
In our simulations $A_\Omega=1.04 \times 10^{-11}$ $\textrm{rad}^2/\textrm{Hz}$, $\omega_h/2\pi=2$ MHz, $\omega_l/2\pi=0.01$ MHz and $A_z=10^8$ Hz$^2$. To examine the impact of the dephasing noise strength on the reconstruction, we vary the scale factor $C$ between $299.1$ and $3.18 $. The resulting $T_2$ times range from $4 \; \mu$s to $100 \; \mu$s. The detuning error $\Delta$ varies between $0.01$ MHz and $0.19$ MHz. 

\begin{figure*}[ht]
\begin{subfloat}{}
    \includegraphics[clip,width=\textwidth]{\detokenize{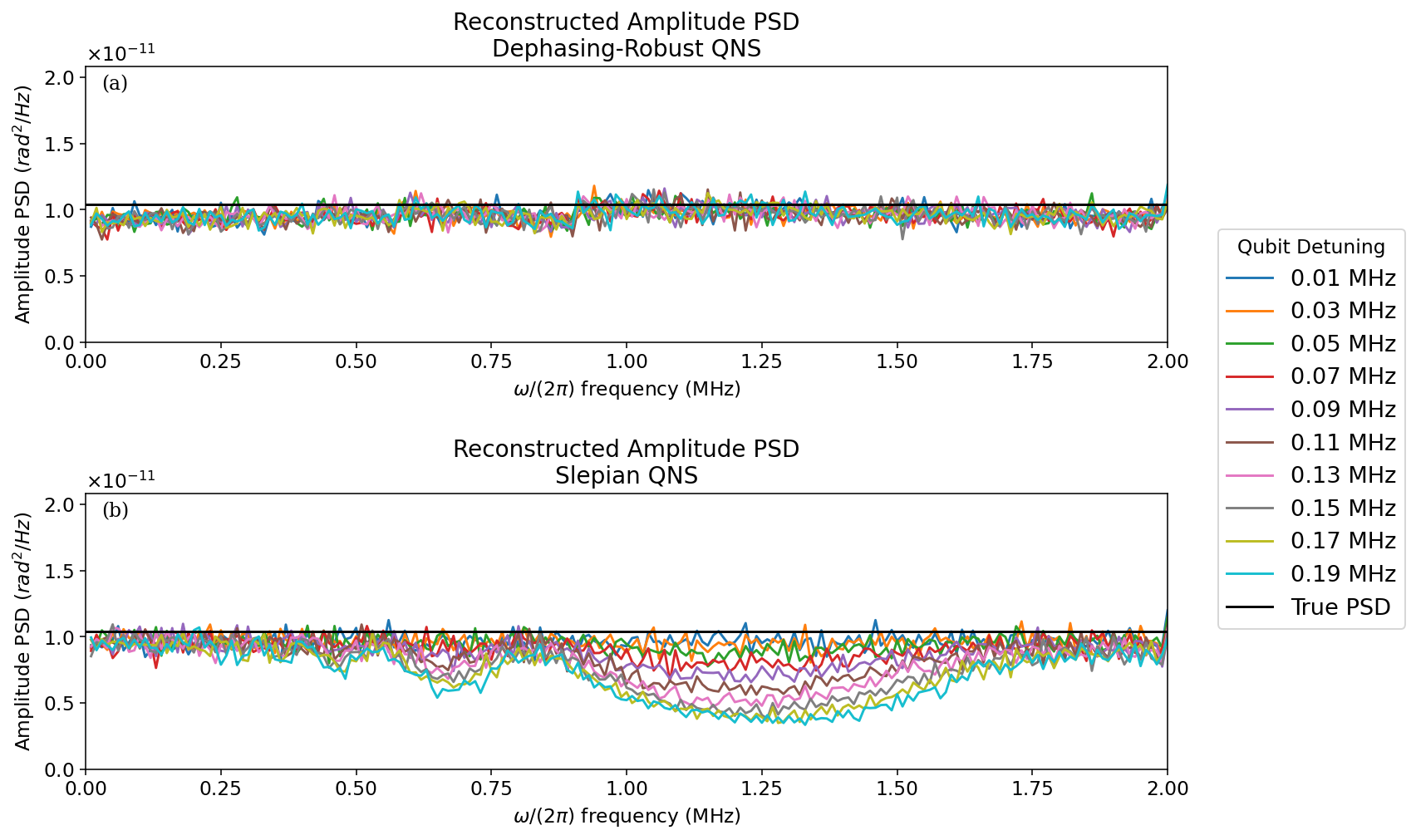}}
    \caption{Amplitude noise QNS with detuning errors of different magnitudes.
    Reconstructions of the amplitude noise spectrum are depicted for the dephasing-robust waveforms (a) and the  DPSS waveforms (b) in the presence of detuning errors ranging from $\Delta=0.01$ MHz to $0.19$ MHz.
    }
    \label{fig:detuning-plots}
\end{subfloat}
\end{figure*}
\begin{figure*}[ht!]
\begin{subfloat}{}
    \includegraphics[width=1.0\textwidth]{\detokenize{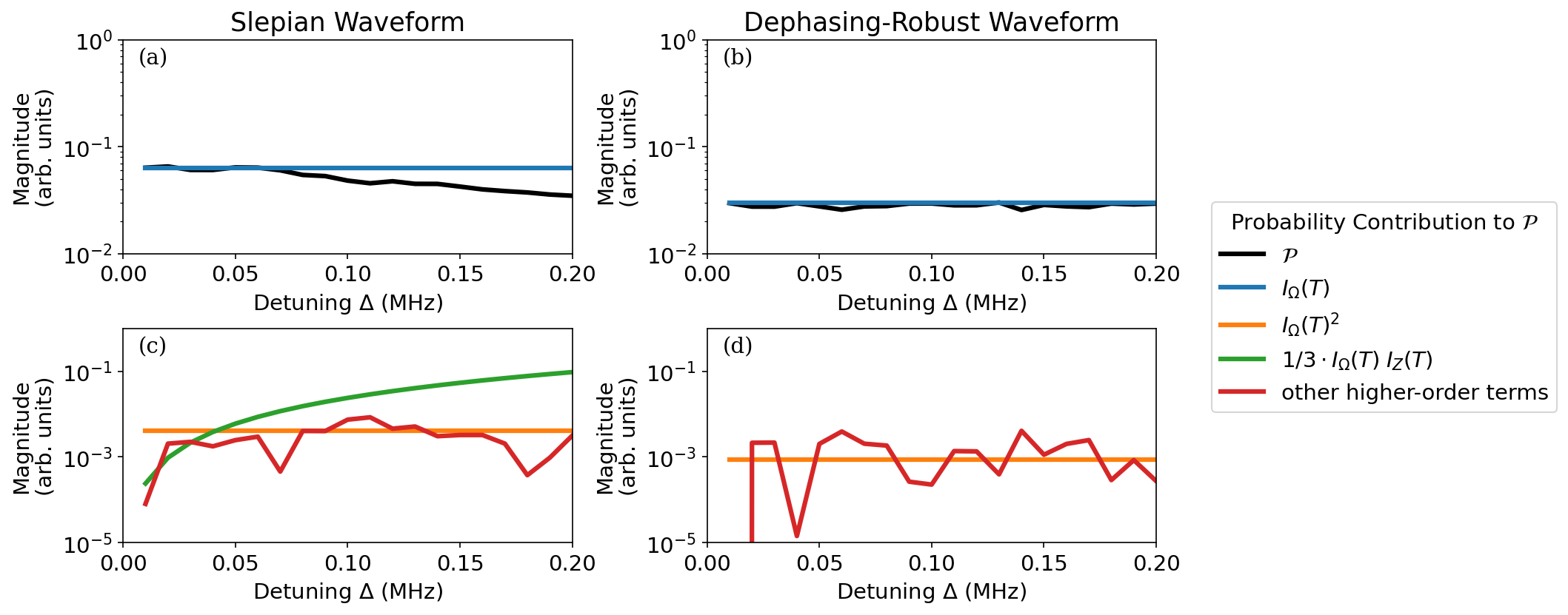}} 
        \caption{Contributions to the survival probability in \eqq{eq::axMeasHigherOrder} with detuning error of various strengths for the DPSS (a,c)  and the dephasing-robust waveforms (b,d). Both the DPSS and dephasing-robust waveforms have modulation frequency $\lambda/2\pi = 1.00$ MHz with all other parameters identical to those used 
        for the reconstructions in Fig.~\ref{fig:detuning-plots}. Plots (a) and (b) show the discrepancy between  $\mathcal{P}\equiv\frac{1}{2} \left [ 1 \!+ \mathbb{P}(\uparrow_1, T) \! - \mathbb{P}(\uparrow_2, T) \! - \mathbb{P}(\uparrow_3, T) \right ]$ and $I_\Omega(T)$ for the DPSS waveform and the dephasing-robust waveform, respectively. The discrepency is determined by the higher-order dephasing-dependent contributions to \eqq{eq::axMeasHigherOrder}, which are plotted in (c) for the DPSS waveform and in (d) for the dephasing-robust waveform. For the dephasing-robust waveform, the $1/3 \cdot I_{\Omega}(T)I_Z(T)$ contribution is exactly cancelled because the DC component of the dephasing FF vanishes, i.e. $F_Z(0,T)=0$.}
        \label{fig:DR-probability-contribution-plots}
\end{subfloat}
\end{figure*}

For control, we employ the dephasing-robust and DPSS waveforms similar to those depicted in Fig.~\ref{fig:big-ff-plots} with an expanded range of modulation frequencies. Concretely, since typical laboratory control hardware generates piece-wise constant waveforms, we discretize the dephasing robust waveform in \eqq{magical-time-domain} into $N$ increments of duration $\Delta t$ so that the resulting amplitude waveform is
\begin{align}\label{eq::DRsimulation}
\Omega(t)=\begin{cases}
0,& t\in[0,\Delta t),\\
\Omega_0\sin(\lambda\Delta t),& t\in[\Delta t,2\Delta t),\\
\;\;\;\;\;\;\;\;\vdots&\;\;\;\;\;\;\;\;\;\;\vdots\\
\Omega_0\sin[\lambda(N-1)\Delta t],& t\in[(N-1)\Delta t,N\Delta t).
\end{cases}
\end{align} 
For the waveform parameters, we use  $N=10,000$, $\Delta t=10$ ns, and $T=N\Delta t=100\,\mu$s. The modulation frequency takes values $\lambda\in\{\Delta\omega,\ldots,L\Delta\omega\}$ with $\Delta\omega/2\pi=0.01$ MHz and $L=200$.
If we use only the dephasing-robust waveforms generated by the first root of $J_0$, we would need to vary the amplitude $\Omega_0$ so that $\Omega_0/\lambda\approx 2.40$ for all $\lambda$. Instead, we use up to the 159th root of $J_0$, using larger roots for smaller $\lambda$, so that $\Omega_0/2\pi\approx 5$ MHz for all modulation frequencies.

For the DPSS, we use waveforms of the lowest ($k=0$) order
combined with sine modulation,
\begin{widetext}
\begin{align}
\Omega(t)=\begin{cases}
0,& t\in[0,\Delta t),\\
\Omega_\text{max}\,v_1^{(0)}(N,W)\sin(\lambda\Delta t),& t\in[\Delta t,2\Delta t),\\
\;\;\;\;\;\;\;\;\;\;\;\;\;\;\;\;\vdots&\;\;\;\;\;\;\;\;\;\;\vdots\\
\Omega_\text{max}\,v_{N-1}^{(0)}(N,W)\sin[\lambda (N-1)\Delta t],& t\in[(N-1)\Delta t,N\Delta t).
\end{cases}\label{eq::DPSSsimulation}
\end{align} 
\end{widetext}
The $k=0$ DPSS, which are the most spectrally concentrated, generate a net identity gate when combined with sine modulation since
\begin{align*}
&v_n^{(0)}(N,W)=v_{N-n}^{(0)}(N,W),\\
&\sin(\lambda n\Delta t)=-\sin[\lambda (N-n)\Delta t],
\end{align*}
ensuring $\int_0^T dt\Omega(t)=0$. In \eqq{eq::DPSSsimulation},  $\Omega_\text{max}/2\pi=5$ MHz, while the other parameters are identical to the dephasing-robust waveform, i.e. $N=10,000$, $\Delta t=10$ ns, $T=N\Delta t=100\,\mu$s, and $\lambda\in\{\Delta\omega,\ldots,L\Delta\omega\}$ with $\Delta\omega/2\pi=0.01$ MHz and $L=200$. We select the bandwidth parameter $W$ of the DPSS so that $NW=1$, which creates an amplitude FF concentrated in the passband $B_\text{DPSS}(\lambda)=[\lambda-2\pi/T,\lambda+2\pi/T]$, equivalent to the passband of a dephasing-robust waveform at modulation frequency $\lambda$. Although our simulations use DPSS and dephasing-robust waveforms with the same passbands, the DPSS waveforms are more spectrally concentrated with a spectral concentration ratio [see Appendix \ref{app:DPSS}] of 0.981. The spectral concentration of the dephasing-robust waveforms is 0.904, for comparison.

To reconstruct the amplitude noise spectrum, we use an inversion approach similar to Ref. \cite{murphy2021universal}. By discretizing the overlap integral in \eqq{eq::overlap} into increments $\Delta\omega$ up to the maximum frequency $L\Delta\omega$, we obtain
\begin{align}
\expect{a_1^{(1)}(T)^2}\approx&\,\frac{1}{\pi}\sum_{\ell=1}^{L}\int_{(\ell-\frac{1}{2} -\frac{1}{2} \delta_{\ell, 1})\Delta\omega}^{(\ell+\frac{1}{2})\Delta\omega}\!\!\!\!d\omega\,F_\Omega(\omega,T)\, S_\Omega(\omega)\notag\\
\approx&\, \vec{F}_\Omega(T)\cdot\vec{S}_\Omega,\label{eq::LinearEq}
\end{align}
where $\vec{F}_\Omega(T)$ and $\vec{S}_\Omega$ are $L$-dimensional vectors with elements
\begin{align}
&[\vec{F}_\Omega(T)]_{\ell=1}=\frac{1}{\pi}\int_{0}^{\frac{3}{2}\Delta\omega}\!\!\!d\omega F_\Omega(\omega,T),\label{eq::FOmega1}\\
&[\vec{F}_\Omega(T)]_{\ell>1}=\frac{1}{\pi}\int_{(\ell-\frac{1}{2}) \Delta\omega}^{(\ell+\frac{1}{2})\Delta\omega}\!\!\!d\omega F_\Omega(\omega,T),\notag\\
&[\vec{S}_\Omega]_\ell=S_\Omega\!\left(\ell \Delta\omega\right)\notag
\end{align}
for $\ell\in\{1,\ldots,L\}$. Note that the integration region defining the first element of $\vec{F}_\Omega$ in \eqq{eq::FOmega1} is larger than the subsequent regions in order to ensure it encloses the peak of the amplitude FF centered at $\lambda =\Delta\omega$.
Equation (\ref{eq::LinearEq}) shows that measuring the $\expect{a_1^{(1)}(T)^2}$ generated by the waveforms at each modulation frequency $\lambda\in\{\Delta\omega,\ldots,L\Delta\omega\}$ produces a linear system that can be inverted to obtain $\vec{S}_\Omega$. Since $S_\Omega(\omega)$ is non-negative, the reconstruction problem is an instance of non-negative least-squares regression, which is solvable through standard numerical techniques. 

\begin{figure*}[ht!]
\begin{subfloat}{}
    \includegraphics[width=1.0\textwidth]{\detokenize{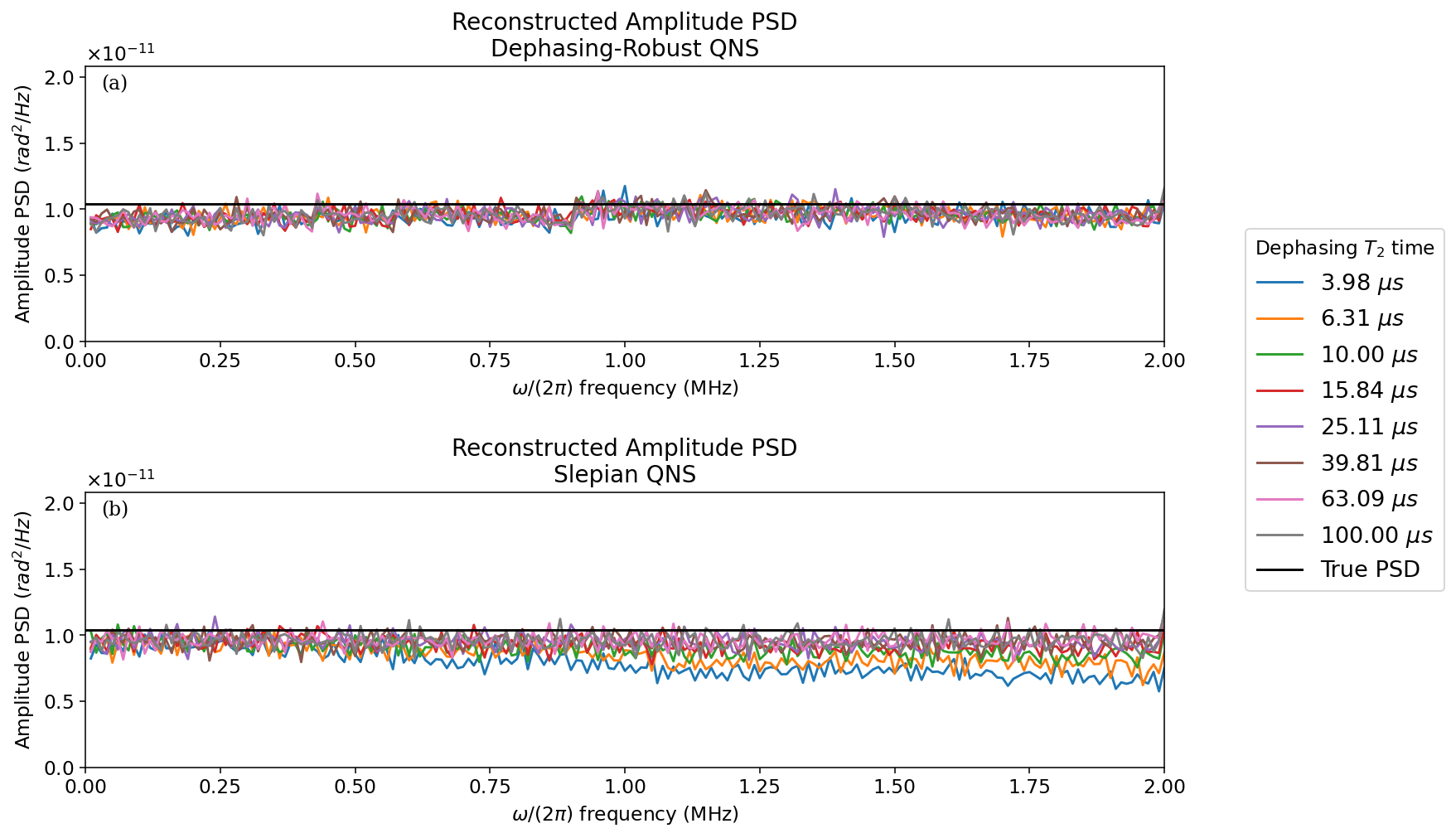}}
    \caption{Amplitude noise QNS with variable strength dephasing.
    Reconstruction of the amplitude noise spectrum using the dephasing-robust waveforms (a) and the  DPSS waveforms (b) in the presence of dephasing noise with $T_2$ times ranging from $4 - 100$ $\mu$s.}
    \label{fig:magical-reconstruction-plots}
\end{subfloat}
\end{figure*}

In Fig.~\ref{fig:detuning-plots}, we use this reconstruction technique to compare the performance of the dephasing-robust and DPSS waveforms in the presence of detuning error of variable strength. The accuracy of the DPSS reconstruction in \ref{fig:detuning-plots}(b) clearly degrades as the strength of the detuning error increases. In contrast, the dephasing-robust reconstruction in \ref{fig:detuning-plots}(a) remains relatively accurate. The difference in performance between the two waveforms is directly related to the higher order terms that bias the tomographic measurement strategy in \eqq{eq:axMeasHigherOrderCompact}, as we demonstrate in Fig.~\ref{fig:DR-probability-contribution-plots}.
Recall that the tomographic measurement strategy approximates the variance $\langle\Delta a_1^{(1)}(T)^2 \rangle$ as a linear combination of survival probabilities, $\mathcal{P}\equiv\frac{1}{2} \left [ 1 \!+ \mathbb{P}(\uparrow_1, T) \! - \mathbb{P}(\uparrow_2, T) \! - \mathbb{P}(\uparrow_3, T) \right ]$. Figure~\ref{fig:DR-probability-contribution-plots}(a) shows that for the DPSS waveforms, $\langle\Delta a_1^{(1)}(T)^2 \rangle$ and $\mathcal{P}$ diverge as the detuning error increases. For the dephasing-robust waveforms, $\langle\Delta a_1^{(1)}(T)^2 \rangle$ and $\mathcal{P}$ are closely matched over the range of detuning errors. This occurs because the dephasing-robust waveforms reduce the magnitudes of the higher-order detuning-dependent terms as compared to the DPSS, as shown in Fig.~\ref{fig:DR-probability-contribution-plots}(c) and \ref{fig:DR-probability-contribution-plots}(d). While the dephasing-robust waveforms successfully mitigate the higher-order detuning-dependent terms, the amplitude noise-dependent term $I_{\Omega}(T)^2$ is one of the few higher-order contributions that remains. This term largely accounts for the slight difference between the reconstruction and true amplitude noise spectrum in Fig.  \ref{fig:detuning-plots}(a).

Figure~\ref{fig:magical-reconstruction-plots} compares the performance of the dephasing-robust and DPSS waveforms in the presence of low-frequency dephasing noise of variable strength. The reconstruction is carried out for dephasing noise with a range of $T_2$ times differing by over an order of magnitude, with smaller $T_2$ indicating stronger noise. Due to bias from low-frequency dephasing, the DPSS reconstruction deviates from the true amplitude noise spectrum for the lowest $T_2$ times. Interestingly, the Slepian reconstruction also exhibits more error at higher modulation frequencies.  Because of their filtering properties, the dephasing-robust waveforms produce an estimate of the amplitude noise spectrum that remains relatively accurate for all strengths of dephasing noise and modulation frequencies. Note, however, that the DPSS reconstruction is only qualitatively affected by the dephasing noise when the $T_2$ time is extremely short relative to the experiment duration of $T=100\;\mu s$. As such, we conclude detuning error is more detrimental to amplitude noise QNS than pure $1/f$ dephasing noise.

While this analysis focuses the relationship between the strength of dephasing noise or detuning error and the higher-order terms that bias amplitude noise QNS, it is important to note that the strength of the amplitude noise plays an important role as well as the interplay between dephasing and detuning. Recall that \eqq{eq:axMeasHigherOrderCompact}, which shows the higher-order corrections to the tomographic measurement strategy, demonstrates a negative multiplicative bias, $I_\Omega(T)I_Z(T)/3$, dependent on $F_Z(\omega,T)$ as well as a positive additive bias dependent on $G_Z(\omega,\omega',T)$. The observation that the DPSS waveforms underestimate the true amplitude noise spectrum in Figs. \ref{fig:detuning-plots}(b) and \ref{fig:magical-reconstruction-plots}(b), as well as the relative size of $I_\Omega(T)I_Z(T)/3$ in  Fig.~\ref{fig:DR-probability-contribution-plots}(c), indicate that the negative bias was the dominant source of error in these simulations.
As the negative bias is proportional to $I_{\Omega}(T)$, we expect it to introduce the same proportionate error for a wide range of amplitude noise strengths. The additive bias due to $G_Z(\omega,\omega,T)$, in contrast, would become more significant for weaker amplitude noise strengths. Additionally, because certain terms contributing to the additive bias in \eqq{eq:GZ} are only nonzero when both dephasing noise and detuning error are present, we may observe different qualitative effects in this scenario. These effects are likely to be much less impactful for the dephasing-robust waveforms because the DC components of $G_Z(\omega,\omega',T)$ are largely suppressed, as shown in Fig.~\ref{fig:Gz}.

\section{Conclusion}
In this work, we used a combination of numerical and analytical methods to determine a family of control waveforms that enable accurate spectroscopy of amplitude control noise in the presence of strong, low-frequency dephasing noise and detuning errors. These waveforms are unique in that they suppress low-frequency dephasing noise and detuning errors while simultaneously producing a spectrally concentrated amplitude FF. To establish the mechanism through which dephasing and detuning can bias estimates of the amplitude noise spectrum, we extended the theoretical framework of Refs. \cite{frey2017NatComm,norris2018optimally} to include both zero-mean noise and higher order perturbative terms. This higher order perturbative analysis enabled us to formulate an objective function quantifying the dynamical contribution of dephasing and cast the task of minimizing dephasing-dependent bias as an optimal control problem. Our numerical optimization utilized techniques based on linear programming to scalably incorporate realistic constraints on control waveforms, such as limitations on the maximum amplitude. The numerically optimized waveforms closely matched the family of analytic dephasing-robust waveforms. Analysis of the primary and higher-order dephasing FFs verified that the dephasing-robust waveforms filter the dynamical contributions of low-frequency dephasing noise and detuning error. 

In numerically simulated amplitude noise QNS experiments, the filtering properties of the dephasing-robust waveforms enabled accurate characterization of amplitude noise in the presence of low-frequency dephasing noise and detuning error. In the case of detuning error, the dephasing-robust robust waveforms offered significant improvement over existing protocols based on DPSS waveforms. For $1/f$ dephasing noise, on the other hand, the performance of dephasing-robust and DPSS waveforms was comparable for all but the lowest $T_2$ times we considered. The performance gains offered by the dephasing-robust waveforms in the presence of a combination of detuning error and dephasing noise, or dephasing noise with a spectrum other than $1/f$, is an area of future study.

Another extension of this work involves our linear programming-inspired optimization procedure. The waveform and amplitude constraints that our method scalably incorporates are relevant to a broad range of laboratory control hardware and more general optimal control problems. Immediate applications include filter design for noise mitigation and QNS in other settings with bounded-strength controls.

\section{Acknowledgements}
It is a pleasure to thank Lorenza Viola for useful discussions, as well as Colin Trout and Kevin Schultz for valuable feedback and critical reading of this manuscript. 
This work was supported by the U.S. Department of Energy, Office of Science, Office of Advanced Scientific Computing Research, Accelerated Research in Quantum Computing under Award Number DE-SC0020316.

\begin{widetext}
\begin{appendix}
\newpage
\section{Measurements}\label{app::Measurements}
\subsection{Survival probabilities}
Consider an experiment in which the qubit is prepared in $\ket{\!\uparrow_i}$ and evolves under the rotating frame Hamiltonian $H(t)=H_N(t)+H_\text{ctrl}(t)$ from $t=0$ to $t=T$. At time $T$, the qubit is measured projectively in the $\ket{\!\uparrow_i},\,\ket{\!\downarrow_i}$ basis, with $\ket{\!\uparrow_i},\,\ket{\!\downarrow_i}$ denoting the $+1,\,-1$ eigenstates of $\sigma_i$, respectively. If this experiment is repeated, the expected fraction of experiments in which $\ket{\!\uparrow_i}$ is measured is given by the survival probability,
\begin{align}\label{eq::SurvivalPAppendix}
\mathbb{P}(\uparrow_i, T) &= \big\langle  \bra{\uparrow_i} {U}(T) \ket{\uparrow_i} ^2  \big\rangle,
\end{align}
where $\expect{\cdot}$ is the ensemble average over realizations of the dephasing and amplitude noise.
 If the amplitude waveform in \eqq{eq::HcAmplitude} generates an identity gate over time $t\in[0,T]$, the rotating frame and toggling frame propagators at time $T$ are equivalent, i.e. $U(T) =U_c(T) \tilde{U}(T) = \tilde{U}(T)$. By substituting $U(T)= \tilde{U}(T)$ in \eqq{eq::SurvivalPAppendix} and using the parametrization of $\tilde{U}(T)$ in terms of the error vector  $\vec{a}(T) \equiv [a_1(T), a_2(T), a_3(T)]$ in \eqq{eq::a}, we obtain
 \begin{align}\label{eq::SurvivalPAppendix2}
\mathbb{P}(\uparrow_i, T) &= \big\langle  \bra{\uparrow_i} \tilde{U}(T) \ket{\uparrow_i} ^2  \big\rangle= \big\langle  \bra{\uparrow_i} e^{-i\vec{a}(T)\cdot\vec{\sigma}} \ket{\uparrow_i} ^2  \big\rangle.
\end{align}

\subsection{Derivation of \eqq{eq::axMeas} with higher order terms}
Evaluating the survival probabilities in \eqq{eq::SurvivalPAppendix2} for $i\in\{1,2,3\}$ and taking a Taylor expansion to fourth order in the error vector components produces 
\begin{align}
\mathbb{P}(\uparrow_1, T)=&
\left\langle\frac{a_2(T)^2+a_3(T)^2+2a_1(T)^2+[a_2(T)^2+a_3(T)^2]\cos\big[2\sqrt{a_1(T)^2+a_2(T)^2+a_3(T)^2}\big]}{2\big[a_1(T)^2+a_2(T)^2+a_3(T)^2\big]}\right\rangle\notag\\\label{eq::P1}
=&\,1-\expect{a_2(T)^2}-\expect{a_3(T)^2}+\frac{1}{3}\big[\expect{a_2(T)^4}
+\expect{a_3(T)^4}+\expect{a_2(T)^2a_1(T)^2}+\expect{a_3(T)^2a_1(T)^2}+2\expect{a_2(T)^2a_3(T)^2}\big],\\
\mathbb{P}(\uparrow_2, T)=&\left\langle
\frac{a_1(T)^2+a_3(T)^2+2a_2(T)^2+[a_1(T)^2+a_3(T)^2]\cos\big[2\sqrt{a_1(T)^2+a_2(T)^2+a_3(T)^2}\big]}{2\big[a_1(T)^2+a_2(T)^2+a_3(T)^2\big]}\right\rangle\notag\\
=&\,1-\expect{a_1(T)^2}-\expect{a_3(T)^2}+\frac{1}{3}\big[\expect{a_1(T)^4}
+\expect{a_3(T)^4}+\expect{a_2(T)^2a_1(T)^2}+\expect{a_3(T)^2a_2(T)^2}+2\expect{a_3(T)^2a_1(T)^2}\big],\label{eq::P2}
\\\mathbb{P}(\uparrow_3, T)=&
\left\langle\frac{a_1(T)^2+a_2(T)^2+2a_3(T)^2+\big[a_1(T)^2+a_2(T)^2\big]\cos\big[2\sqrt{a_1(T)^2+a_2(T)^2+a_3(T)^2}\big]}{2\big[a_1(T)^2+a_2(T)^2+a_3(T)^2\big]}\right\rangle\notag\\
=&\,1-\expect{a_2(T)^2}-\expect{a_1(T)^2}+\frac{1}{3}\big[\expect{a_2(T)^4}
+\expect{a_1(T)^4}+\expect{a_2(T)^2a_3(T)^2}+\expect{a_3(T)^2a_1(T)^2}+2\expect{a_1(T)^2a_2(T)^2}\big].\label{eq::P3}
\end{align}

From the Magnus expansion of the error propagator, $\tilde{U}(T)=e^{-i\sum_{m=1}^\infty M^{(m)}(T)}$, we can determine a  perturbative expansion of the error vector by projecting the terms of the Magnus expansion into the Pauli basis so that $\vec{a}=\sum_{m=1}^\infty \vec{a}^{(m)}(T)$, where $a_i^{(m)}(T)=\text{Tr}[\sigma_i M^{(m)}(T)]\sim T^m$ for $i\in\{1,2,3\}$. The first two terms in the error vector expansion have components
\begin{align}
&a_i^{(1)}(T)=\frac{1}{2}\text{Tr}\left\{\sigma_i\int_0^T\!\!\!dt\,\tilde{H}(t)\right\},\label{eq::a1Expansion}\\
&a_i^{(2)}(T)=-\frac{i}{4}\text{Tr}\left\{\sigma_i\int_0^T\!\!\!dt\int_0^t\!\!\!dt'\,\big[\tilde{H}(t),\tilde{H}(t')\big]\right\}.
\label{eq::a2Expansion}
\end{align}
By taking $a_i(T)=a_i^{(1)}(T)+a_i^{(2)}(T)+\ldots$ and keeping terms up to order $T^4$ in Eqs. (\ref{eq::P1})-(\ref{eq::P3}), we find
\begin{align}\notag
&\frac{1}{2}\left[1+\mathbb{P}(\uparrow_1, T)-\mathbb{P}(\uparrow_2, T)-\mathbb{P}(\uparrow_3, T)\right]\\&=\langle a_1^{(1)}(T)^2 \rangle+2\expect{a_1^{(1)}(T) a_1^{(2)}(T)}+
\langle a_1^{(2)}(T)^2 \rangle\!
-\frac{1}{3}\big[\langle a_1^{(1)}(T)^4 \rangle
  +\langle a_2^{(1)}(T)^2\,a_1^{(1)}(T)^2\rangle
  +\langle a_3^{(1)}(T)^2\,a_1^{(1)}(T)^2\rangle\big]\notag\\
  &=\langle \Delta a_1^{(1)}(T)^2\rangle\!+\!\expect{a_1^{(1)}(T)}^2 \!+\!2\expect{a_1^{(1)}(T) a_1^{(2)}(T)}\!+\!
\langle a_1^{(2)}(T)^2 \rangle\!
-\frac{1}{3}\big[\langle a_1^{(1)}(T)^4 \rangle
  +\langle a_2^{(1)}(T)^2\,a_1^{(1)}(T)^2\rangle
  +\langle a_3^{(1)}(T)^2\,a_1^{(1)}(T)^2\rangle\big].
\label{eq::HigherOrderTerms}
\end{align}
Because they contain odd moments of $\beta_\Omega(t)$, the second and third terms vanish when the amplitude noise is zero-mean. In this case, we recover \eqq{eq::axMeasHigherOrder} from the main text.

\subsection{Explicit expressions of the terms in \eqq{eq::HigherOrderTerms}}
The second term in \eqq{eq::HigherOrderTerms}  arises when the amplitude noise has a nonzero mean. Its form follows in a straightforward manner from \eqq{eq::a1},
\begin{align}\notag
\expect{a_1^{(1)}(T)}^2=&\,\frac{1}{4}\left[\int_{0}^{T} \!\!\mathrm{d}t \,  \Omega (t)\, \expect{\beta_{\Omega}(t)}\right]^2
\\=&\,\frac{\mu_\Omega^2}{4}\left[\int_{0}^{T} \!\!\mathrm{d}t \,  \Omega (t)\right]^2.\label{eq::meanterm1}
\end{align}
Because the mean of a stationary process is time-independent, we have taken $\mu_\Omega\equiv\expect{\beta_{\Omega}(t)}$ in the last line.

The third term in \eqq{eq::HigherOrderTerms}, which depends on a third order moment of $\beta_\Omega(t)$ and $\beta_z(t)$, is also non-vanishing when the amplitude noise has nonzero mean. Using the Gaussianity and independence of $\beta_\Omega(t)$ and $\beta_z(t)$, we can factorize the third order moment into a product of one- and two-point correlation functions,
\begin{align*}
\expect{\beta_\Omega(t_1)\beta_z(t_2)\beta_z(t_3)}=&\,\expect{\beta_\Omega(t_1)\beta_z(t_2)}\expect{\beta_z(t_3)}+\expect{\beta_\Omega(t_1)\beta_z(t_3)}\expect{\beta_z(t_2)}+\expect{\beta_z(t_2)\beta_z(t_3)}\expect{\beta_\Omega(t_1)}\\&-2\expect{\beta_\Omega(t_1)}\expect{\beta_z(t_2)}\expect{\beta_z(t_3)}\\=&\,\expect{\beta_z(t_2)\beta_z(t_3)}\expect{\beta_\Omega(t_1)}.
\end{align*}
Using this expression and \eqq{eq::a2Expansion} to evaluate $a_1^{(2)}(T)$, we find
\begin{align}
\expect{a_1^{(1)}(T) a_1^{(2)}(T)}=\frac{\mu_\Omega}{2}\left[\int_{0}^{T} \!\!\mathrm{d}t_1 \,  \Omega (t_1)\right]\int_0^T\!\!\!dt_2\int_0^{t_2}\!\!\!dt_3\,\sin\big[\Theta(t_2)-\Theta(t_3)\big]\expect{\beta_z(t_2)\beta_z(t_3)}.\label{eq::meanterm2}
\end{align}

The remaining higher-order terms in \eqq{eq::HigherOrderTerms} depend on fourth-order moments of $\beta_\Omega(t)$ and $\beta_z(t)$. Like the third-order moment above, the fourth-order moments factor into products of one- and two-point correlation functions. For $i_1,i_2,i_3,i_4\in\{\Omega,z\}$, an arbitrary fourth moment can be written as
\begin{align}\label{eq::4thMoment}
\big\langle\beta_{i_1}(t_1)\beta_{i_2}(t_2)\beta_{i_3}(t_3)\beta_{i_4}(t_4)\big\rangle=&\,
\big\langle\beta_{i_1}(t_1)\beta_{i_2}(t_2)\big\rangle
\big\langle\beta_{i_3}(t_3)\beta_{i_4}(t_4)\big\rangle+
\big\langle\beta_{i_1}(t_1)\beta_{i_3}(t_3)\big\rangle
\big\langle\beta_{i_2}(t_2)\beta_{i_4}(t_4)\big\rangle\\&+
\big\langle\beta_{i_1}(t_1)\beta_{i_4}(t_4)\big\rangle
\big\langle\beta_{i_3}(t_3)\beta_{i_2}(t_2)\big\rangle-
2\expect{\beta_{i_1}(t_1)}\expect{\beta_{i_2}(t_2)}\expect{\beta_{i_3}(t_3)}\expect{\beta_{i_4}(t_4)}.\notag
\end{align}
From this expression, we can determine the higher order terms in the frequency domain. 

Consider the fifth, sixth and seventh terms on the RHS of \eqq{eq::HigherOrderTerms}, which depend on $\vec{a}^{(1)}(T)$ and can be written as
\begin{align}\label{eq::a1Terms}
\langle a_i^{(1)}(T)^2\, a_j^{(1)}(T)^2\rangle
=&\int_0^T\!\!\!dt_1 \!\!\int_0^T\!\!\! dt_2\!\!\int_0^T\!\!\! dt_3\!\!\int_0^T\!\!\! dt_4\; y_i(t_1)y_i(t_2)y_j(t_3)y_j(t_4)\,\Big[\big\langle\beta_{i}(t_1)\beta_{i}(t_2)\big\rangle
\big\langle\beta_{j}(t_3)\beta_{j}(t_4)\big\rangle\\\notag&+\big\langle\beta_{i}(t_1)\beta_{j}(t_3)\big\rangle
\big\langle\beta_{i}(t_2)\beta_{j}(t_4)\big\rangle
+\big\langle\beta_{i}(t_1)\beta_{j}(t_4)\big\rangle
\big\langle\beta_{i}(t_2)\beta_{j}(t_3)\big\rangle\\\notag&-2\big\langle\beta_{i}(t_1)\big\rangle\big\langle\beta_{i}(t_2)\big\rangle\big\langle\beta_{j}(t_3)\big\rangle\big\langle\beta_{j}(t_4)\big\rangle\Big],
\end{align}
where $i\in\{1,2,3\}$ and $j=1$.

When $i\neq j$, the independence of $\beta_\Omega(t)$ and $\beta_z(t)$ implies that the last three terms cancel, leaving 
\begin{align}
\langle a_i^{(1)}(T)^2\, a_j^{(1)}(T)^2\rangle_{i\neq j}
=&\int_0^T\!\!\!dt_1 \!\!\int_0^T\!\!\! dt_2\!\!\int_0^T\!\!\! dt_3\!\!\int_0^T\!\!\! dt_4\; y_i(t_1)y_i(t_2)y_j(t_3)y_j(t_4)\,\big\langle\beta_{i}(t_1)\beta_{i}(t_2)\big\rangle
\big\langle\beta_{j}(t_3)\beta_{j}(t_4)\big\rangle\\\notag
=&\int_0^T\!\!\!dt_1 \!\!\int_0^T\!\!\! dt_2\!\!\int_0^T\!\!\! dt_3\!\!\int_0^T\!\!\! dt_4\; y_i(t_1)y_i(t_2)y_j(t_3)y_j(t_4)\,\Big[\big\langle\Delta\beta_{i}(t_1)\Delta\beta_{i}(t_2)\big\rangle
\big\langle\Delta\beta_{j}(t_3)\Delta\beta_{j}(t_4)\big\rangle\\
\notag&+\big\langle\Delta\beta_{i}(t_1)\Delta\beta_{i}(t_2)\big\rangle\big\langle\beta_{j}(t_3)\big\rangle \big\langle\beta_{j}(t_4)\big\rangle+\big\langle\beta_{i}(t_1)\big\rangle \big\langle\beta_{i}(t_2)\big\rangle
\big\langle\Delta\beta_{j}(t_3)\Delta\beta_{j}(t_4)\big\rangle\\
&+\big\langle\beta_{i}(t_1)\big\rangle \big\langle\beta_{i}(t_2)\big\rangle\big\langle\beta_{j}(t_3)\big\rangle \big\langle\beta_{j}(t_4)\big\rangle\Big]
\end{align}
When $\beta_j(t)=\beta_\Omega(t)$ is zero-mean, this expression reduces to
\begin{align*}
\langle a_i^{(1)}(T)^2\, a_j^{(1)}(T)^2\rangle_{i\neq j}=\int_0^T\!\!\!dt_1 \!\!\int_0^T\!\!\! dt_2\!\!\int_0^T\!\!\! dt_3\!\!\int_0^T\!\!\! dt_4\; y_i(t_1)y_i(t_2)y_j(t_3)y_j(t_4)\,\Big[&\big\langle\Delta\beta_{i}(t_1)\Delta\beta_{i}(t_2)\big\rangle
+\big\langle\beta_{i}(t_1)\big\rangle \big\langle\beta_{i}(t_2)\big\rangle\Big]\\&\times\big\langle\Delta\beta_{j}(t_3)\Delta\beta_{j}(t_4)\big\rangle
\end{align*}
From Eqs. (\ref{eq::a1})-(\ref{eq::a3}) and (\ref{eq::beta_vec}), we then find 
\begin{align}\notag
\langle a_1^{(1)}(T)^2&\, a_2^{(1)}(T)^2\rangle
  +\langle a_1^{(1)}(T)^2\, a_3^{(1)}(T)^2\rangle
  \\\notag=&\int_0^T\!\!\!dt_1 \!\!\int_0^T\!\!\! dt_2\Omega(t_1)\Omega(t_2)\big\langle\Delta\beta_\Omega(t_1)\Delta\beta_\Omega(t_2)\big\rangle\int_0^T\!\!\!dt_3 \!\!\int_0^T\!\!\!dt_4 \sin\Theta(t_3)\sin\Theta(t_4)\Big[\big\langle\Delta\beta_z(t_3)\Delta\beta_z(t_4)\big\rangle+\mu_z^2\Big]\\\notag
  &+\int_0^T\!\!\!dt_1 \!\!\int_0^T\!\!\! dt_2\Omega(t_1)\Omega(t_2)\big\langle\Delta\beta_\Omega(t_1)\Delta\beta_\Omega(t_2)\big\rangle\int_0^T\!\!\!dt_3 \!\!\int_0^T\!\!\!dt_4 \cos\Theta(t_3)\cos\Theta(t_4)\Big[\big\langle\Delta\beta_z(t_3)\Delta\beta_z(t_4)\big\rangle+\mu_z^2\Big]
  \\\notag=&\,\bigg\{\frac{1}{2\pi}\!\!\int_{-\infty}^\infty\!\!\! d\omega\,S_\Omega(\omega)F_\Omega(\omega,T)\bigg\}\bigg\{\frac{1}{2\pi}\!\!\int_{-\infty}^\infty\!\!\! d\omega\bigg[S_z(\omega)+2\pi\mu_z^2\delta(\omega)\bigg]\bigg[\;\bigg|\int_0^T\!\!\!dt e^{i\omega t}\sin\Theta(t)\bigg|^2+\bigg|\int_0^T\!\!\!dt e^{i\omega t}\cos\Theta(t)\bigg|^2\;\bigg]\bigg\}\\\notag
  =&\,\bigg\{\frac{1}{2\pi}\int_{-\infty}^\infty\!\!\! d\omega\,S_\Omega(\omega)F_\Omega(\omega,T)\bigg\}\bigg\{\frac{1}{2\pi}\int_{-\infty}^\infty\!\!\! d\omega\bigg[S_z(\omega)+2\pi\mu_z^2\delta(\omega)\bigg]F_Z(\omega,T)\bigg\}
  \\=&\,I_\Omega(T)I_Z(T),\label{eq::IOmIz}
\end{align}
as given in the main text.

When $i=j=1$, on the other hand, we have
\begin{align}\notag
&\langle a_1^{(1)}(T)^4 \rangle=\int_0^T\!\!\!dt_1 \!\!\int_0^T\!\!\! dt_2\!\!\int_0^T\!\!\! dt_3\!\!\int_0^T\!\!\! dt_4\; \Omega(t_1)\Omega(t_2)\Omega(t_3)\Omega(t_4)\,\Big[\big\langle\beta_\Omega(t_1)\beta_\Omega(t_2)\big\rangle
\big\langle\beta_\Omega(t_3)\beta_\Omega(t_4)\big\rangle\\&+
\big\langle\beta_\Omega(t_1)\beta_\Omega(t_3)\big\rangle
\big\langle\beta_\Omega(t_2)\beta_\Omega(t_4)\big\rangle+
\big\langle\beta_\Omega(t_1)\beta_\Omega(t_4)\big\rangle
\big\langle\beta_\Omega(t_3)\beta_\Omega(t_2)\big\rangle-
2\mu_\Omega^4\Big]\notag\\\notag\\\notag
&=\int_0^T\!\!\!dt_1 \!\!\int_0^T\!\!\! dt_2\!\!\int_0^T\!\!\! dt_3\!\!\int_0^T\!\!\! dt_4\; \Omega(t_1)\Omega(t_2)\Omega(t_3)\Omega(t_4)\,
\Big\{\,\big\langle\Delta\beta_\Omega(t_1)\Delta\beta_\Omega(t_2)\big\rangle\big\langle\Delta\beta_\Omega(t_3)\Delta\beta_\Omega(t_4)\big\rangle\\&+\big\langle\Delta\beta_\Omega(t_1)\Delta\beta_\Omega(t_3)\big\rangle\big\langle\Delta\beta_\Omega(t_2)\Delta\beta_\Omega(t_4)\big\rangle+\big\langle\Delta\beta_\Omega(t_1)\Delta\beta_\Omega(t_4)\big\rangle\big\langle\Delta\beta_\Omega(t_2)\Delta\beta_\Omega(t_3)\big\rangle\notag\\&
+\big[\big\langle\beta_\Omega(t_1)\beta_\Omega(t_2)\big\rangle\!+\!\big\langle\beta_\Omega(t_1)\beta_\Omega(t_3)\big\rangle\!+\!\big\langle\beta_\Omega(t_1)\beta_\Omega(4)\big\rangle\!+\!\big\langle\beta_\Omega(t_2)\beta_\Omega(t_3)\big\rangle\!+\!\big\langle\beta_\Omega(t_2)\beta_\Omega(t_4)\big\rangle\!+\!\big\langle\beta_\Omega(t_3)\beta_\Omega(t_4)\big\rangle\big]\mu_\Omega^2-5\mu_\Omega^4\Big\}.\label{eq::meanterm3}
\end{align}
If the amplitude noise is zero-mean, the last line in this expression vanishes, producing
\begin{align*}
\langle a_1^{(1)}(T)^4 \rangle=&\int_0^T\!\!\!dt_1 \!\!\int_0^T\!\!\! dt_2\!\!\int_0^T\!\!\! dt_3\!\!\int_0^T\!\!\! dt_4\; \Omega(t_1)\Omega(t_2)\Omega(t_3)\Omega(t_4)\,\Big[\big\langle\Delta\beta_\Omega(t_1)\Delta\beta_\Omega(t_2)\big\rangle
\big\langle\Delta\beta_\Omega(t_3)\Delta\beta_\Omega(t_4)\big\rangle\\&+
\big\langle\Delta\beta_\Omega(t_1)\Delta\beta_\Omega(t_3)\big\rangle
\big\langle\Delta\beta_\Omega(t_2)\Delta\beta_\Omega(t_4)\big\rangle+
\big\langle\Delta\beta_\Omega(t_1)\Delta\beta_\Omega(t_4)\big\rangle
\big\langle\Delta\beta_\Omega(t_3)\Delta\beta_\Omega(t_2)\big\rangle\Big]\\
=&\,3\!\int_0^T\!\!\!dt_1 \!\!\int_0^T\!\!\! dt_2\Omega(t_1)\Omega(t_2)\big\langle\Delta\beta_\Omega(t_1)\Delta\beta_\Omega(t_2)\big\rangle\int_0^T\!\!\!dt_3 \!\!\int_0^T\!\!\! dt_4\Omega(t_3)\Omega(t_4)\big\langle\Delta\beta_\Omega(t_3)\Delta\beta_\Omega(t_4)\big\rangle\\
=&\,3\left[\frac{1}{2\pi}\int_{-\infty}^\infty\!\!\! d\omega S_\Omega(\omega)F_\Omega(\omega,T)\right]^2\\
=&\,3\,I_\Omega(T)^2,
\end{align*}
as given in the main text.

Next, we consider the fourth term on the RHS of \eqq{eq::HigherOrderTerms}, which depends on $a_1^{(2)}(T)$. Using \eqq{eq::4thMoment} to evaluate the higher order moment and taking $\expect{\beta_z(t)}=\mu_z$ for stationary noise, we obtain
\begin{align}\notag
&\big\langle a_1^{(2)}(T)^2 \big\rangle    =\int_0^T\!\!\!dt_1\int_0^{t_1}\!\!\!dt_2\,\sin\big[\Theta(t_1)-\Theta(t_2)\big]\!\!\int_0^T\!\!\!dt_3\int_0^{t_3}\!\!\!dt_4\,\sin\big[\Theta(t_3)-\Theta(t_4)\big]\,\big\langle\beta_z(t_1)\beta_z(t_2)\beta_z(t_3)\beta_z(t_4)\big\rangle\\\notag\\\notag
&\;\;=\int_0^T\!\!\!dt_1\int_0^{t_1}\!\!\!dt_2\,\sin\big[\Theta(t_1)-\Theta(t_2)\big]\!\!\int_0^T\!\!\!dt_3\int_0^{t_3}\!\!\!dt_4\,\sin\big[\Theta(t_3)-\Theta(t_4)\big]\,\big[\,\big\langle\beta_z(t_1)\beta_z(t_2)\big\rangle\big\langle\beta_z(t_3)\beta_z(t_4)\big\rangle\\&\notag\;\;\;\;\;\;+\big\langle\beta_z(t_1)\beta_z(t_3)\big\rangle\big\langle\beta_z(t_2)\beta_z(t_4)\big\rangle+\big\langle\beta_z(t_1)\beta_z(t_4)\big\rangle\big\langle\beta_z(t_2)\beta_z(t_3)\big\rangle-2\mu_z^4\big]\\\notag\\\notag
&\;\;=\int_0^T\!\!\!dt_1\int_0^{t_1}\!\!\!dt_2\,\sin\big[\Theta(t_1)-\Theta(t_2)\big]\!\!\int_0^T\!\!\!dt_3\int_0^{t_3}\!\!\!dt_4\,\sin\big[\Theta(t_3)-\Theta(t_4)\big]\,\Big\{\,\big\langle\Delta\beta_z(t_1)\Delta\beta_z(t_2)\big\rangle\big\langle\Delta\beta_z(t_3)\Delta\beta_z(t_4)\big\rangle\\\notag&+\big\langle\Delta\beta_z(t_1)\Delta\beta_z(t_3)\big\rangle\big\langle\Delta\beta_z(t_2)\Delta\beta_z(t_4)\big\rangle+\big\langle\Delta\beta_z(t_1)\Delta\beta_z(t_4)\big\rangle\big\langle\Delta\beta_z(t_2)\Delta\beta_z(t_3)\big\rangle\\&
+\big[\big\langle\beta_z(t_1)\beta_z(t_2)\big\rangle+\big\langle\beta_z(t_1)\beta_z(t_3)\big\rangle+\big\langle\beta_z(t_1)\beta_z(4)\big\rangle+\big\langle\beta_z(t_2)\beta_z(t_3)\big\rangle+\big\langle\beta_z(t_2)\beta_z(t_4)\big\rangle+\big\langle\beta_z(t_3)\beta_z(t_4)\big\rangle\big]\mu_z^2-5\mu_z^4\Big\}.\label{eq::meanterm4}
\end{align}
If the dephasing noise is zero-mean, the final line vanishes and we have
\begin{align*}
\big\langle a_1^{(2)}(T)^2 \big\rangle_{\mu_z=0} =&\,\frac{1}{(2\pi)^2}\int_0^T\!\!\!dt_1\int_0^{t_1}\!\!\!dt_2\,\sin\big[\Theta(t_1)-\Theta(t_2)\big]\!\!\int_0^T\!\!\!dt_3\int_0^{t_3}\!\!\!dt_4\,\sin\big[\Theta(t_3)-\Theta(t_4)\big]\\&\;\;\;\bigg[\int_{-\infty}^\infty\!\!\!d\omega\,e^{i\omega(t_1-t_2)}S(\omega)\!\!\int_{-\infty}^\infty\!\!\!d\omega'\,e^{i\omega'(t_3-t_4)}S(\omega')
+\int_{-\infty}^\infty\!\!\!d\omega\,e^{i\omega(t_1-t_3)}S(\omega)\!\!\int_{-\infty}^\infty\!\!\!d\omega'\,e^{i\omega'(t_2-t_4)}S(\omega')
\\&\;\;\;\;\;\;\;\;\;\;\;\;+\int_{-\infty}^\infty\!\!\!d\omega\,e^{i\omega(t_1-t_4)}S(\omega)\!\!\int_{-\infty}^\infty\!\!\!d\omega'\,e^{i\omega'(t_2-t_3)}S(\omega')\bigg].
\end{align*}
Collecting like terms produces
\begin{align*}
\big\langle a_1^{(2)}(T)^2 \big\rangle_{\mu_z=0} =\frac{1}{(2\pi)^2}\int_{-\infty}^\infty\!\!\!d\omega\!\int_{-\infty}^\infty\!\!\!d\omega'\, G_Z(\omega,\omega',T)\,S_z(\omega)S_z(\omega'),
\end{align*}
where 
\begin{align*}
G_Z(\omega,\omega',T)=\int_0^{T}\!\!\!\!dt_1\!\!\int_0^{t_1}\!\!\!\!dt_2\,&\sin\big[\Theta(t_1)\!-\!\Theta(t_2)\big]\int_0^{T}\!\!\!\!dt_3\!\!\int_0^{t_3}\!\!\!\!dt_4\,\sin\big[\Theta(t_3)\!-\!\Theta(t_4)\big]\notag\big[e^{i\omega(t_1-t_2)}e^{i\omega'(t_3-t_4)}
\\&+e^{i\omega(t_1-t_3)}e^{i\omega'(t_2-t_4)}
+e^{i\omega(t_1-t_4)}e^{i\omega'(t_2-t_3)}\big].\notag
\end{align*}
When the dephasing noise is nonzero-mean, the terms in the final line of \eqq{eq::meanterm4} make a contribution,
\begin{align}\notag
\big\langle a_1^{(2)}(T)^2 \big\rangle_{\mu_z\neq 0}=&\,\mu_z^2\int_0^T\!\!\!dt_1\int_0^{t_1}\!\!\!dt_2\,\sin\big[\Theta(t_1)-\Theta(t_2)\big]\!\!\int_0^T\!\!\!dt_3\int_0^{t_3}\!\!\!dt_4\,\sin\big[\Theta(t_3)-\Theta(t_4)\big]\,
\big[\big\langle\beta_z(t_1)\beta_z(t_2)\big\rangle\\\notag&+\big\langle\beta_z(t_3)\beta_z(t_4)\big\rangle
+\big\langle\beta_z(t_1)\beta_z(t_3)\big\rangle+
\big\langle\beta_z(t_2)\beta_z(t_4)\big\rangle
+\big\langle\beta_z(t_1)\beta_z(t_4)\big\rangle+
\big\langle\beta_z(t_2)\beta_z(t_3)\big\rangle-5\mu_z^2\big]\\\notag\\\notag
=&\,\mu_z^2\int_0^T\!\!\!dt_1\int_0^{t_1}\!\!\!dt_2\,\sin\big[\Theta(t_1)-\Theta(t_2)\big]\!\!\int_0^T\!\!\!dt_3\int_0^{t_3}\!\!\!dt_4\,\sin\big[\Theta(t_3)-\Theta(t_4)\big]\,
\big[\big\langle\Delta\beta_z(t_1)\Delta\beta_z(t_2)\big\rangle\\\notag&+\big\langle\Delta\beta_z(t_3)\Delta\beta_z(t_4)\big\rangle
+\big\langle\Delta\beta_z(t_1)\Delta\beta_z(t_3)\big\rangle+
\big\langle\Delta\beta_z(t_2)\Delta\beta_z(t_4)\big\rangle
\\\notag&+\big\langle\Delta\beta_z(t_1)\Delta\beta_z(t_4)\big\rangle+
\big\langle\Delta\beta_z(t_2)\Delta\beta_z(t_3)\big\rangle+\mu_z^2\big]\\\notag\\\notag
=&\,\frac{\mu_z^2}{2\pi}\int_{-\infty}^\infty\!\!\!\!d\omega\int_{-\infty}^\infty\!\!\!\!d\omega'\int_0^{T}\!\!\!\!dt_1\!\!\int_0^{t_1}\!\!\!\!dt_2\,\sin\big[\Theta(t_1)\!-\!\Theta(t_2)\big]\int_0^{T}\!\!\!\!dt_3\!\!\int_0^{t_3}\!\!\!\!dt_4\,\sin\big[\Theta(t_3)\!-\!\Theta(t_4)\big]\notag\Big[e^{i\omega(t_1-t_2)}e^{i\omega'(t_3-t_4)}
\\\notag&+e^{i\omega(t_1-t_3)}e^{i\omega'(t_2-t_4)}
+e^{i\omega(t_1-t_4)}e^{i\omega'(t_2-t_3)}\Big]\Big[S_z(\omega)\delta(\omega')+S_z(\omega')\delta(\omega)
+\frac{2\pi\mu_z^2}{3}\delta(\omega)\delta(\omega')\Big]\\\notag\\
=&\,\int_{-\infty}^\infty\!\!\!d\omega\!\int_{-\infty}^\infty\!\!\!d\omega'\, G_Z(\omega,\omega',T)\Big[\frac{\mu_z^2}{2\pi}S_z(\omega)\delta(\omega')+\frac{\mu_z^2}{2\pi}S_z(\omega')\delta(\omega)
+\frac{\mu_z^4}{3}\delta(\omega)\delta(\omega')\Big].\label{app::NoMeanZ}
\end{align}
By taking $\big\langle a_1^{(2)}(T)^2 \big\rangle_{\mu_z= 0}+\big\langle a_1^{(2)}(T)^2 \big\rangle_{\mu_z\neq 0}$, we recover \eqq{eq::a22} in the main text.

\section{Slepian (DPSS) sequences and spectral concentration}\label{app:DPSS}
For a detailed treatment of the Slepian or discrete prolate spheroidal sequences (DPSS) and their applications in classical signal processing and spectral estimation, see Refs. \cite{dpss,thomson1982spectrum,PandW}. We briefly summarize material from these references here.
The DPSS $\{v_n^{(k)}\}\equiv\{v_0^{(k)},\ldots,v_N^{(k)}\}$ are a family of $N$, length-$N$ sequences that are solutions to a $N\times N$ Toeplitz matrix eigenvalue equation,
\begin{align}\label{eq::DPSSdef}
\sum_{m=0}^{N-1}\frac{\sin 2\pi W(n-m)}{\pi(n-m)}v_m^{(k)}
=\lambda_k(N,W)\,v_n^{(k)}.
\end{align}
Here, $k\in\{0,\ldots,N-1\}$ indexes the order of the DPSS and $W<1/2$ is a bandwidth parameter related to spectral concentration in the frequency domain. The frequency domain representations of the DPSS, known as the discrete prolate spheroidal wave functions (DPSWF), depend on the discrete-time Fourier transform (DTFT) of the DPSS, 
\begin{align*}
U^{(k)}(N,W;\omega)=\epsilon_k\sum_{n=0}^{N-1} v_n^{(k)} e^{i\omega[n-(N-1)/2]\Delta t}.
\end{align*}
Here, $\epsilon_k=1\,(i)$ for $k$ even (odd).

For a general length-$N$ sequence, $\{a_1,\ldots,a_N\}\equiv\{a_n\}$, with DTFT
$\tilde{a}(\omega)=\sum_{n=0}^{N-1}a_n e^{i\omega[n-(N-1)/2]\Delta t}$, spectral concentration in a frequency band $B(0)\equiv[-\delta\omega,\delta\omega]$ is quantified by the ratio
\begin{align*}
\mathcal{C}_B[\{a_n\}]=\frac{\int_{-\delta\omega}^{\delta\omega}d\omega\,|\tilde{a}(\omega)|^2 }{\int_{-\omega_N}^{\omega_N}d\omega\,|\tilde{a}(\omega)|^2}, 
\end{align*}
where $\omega_N\equiv\pi/\Delta t$ is the Nyquist frequency and $\delta\omega<\omega_N$. For $W=\delta\omega\Delta t/2\pi$, the spectral concentration of the $k$th order DPSS in the band $B(0)$ is given by
\begin{align*}
\mathcal{C}_B[\{v_n^{(k)}\}]=\frac{\int_{-\delta\omega}^{\delta\omega}d\omega\,|U^{(k)}(N,W;\omega)|^2 }{\int_{-\omega_N}^{\omega_N}d\omega\,|U^{(k)}(N,W;\omega)|^2}=\lambda_k(N,W), 
\end{align*}
where $\lambda_k(N,W)$ is the eigenvalue from \eqq{eq::DPSSdef}. Of all length-$N$ sequences, the DPSS of order $k=0$ with eigenvalue $\lambda_0(N,W)\approx 1$ has the maximum spectral concentration in $B(0)$. The first $\lfloor NW\rfloor$ DPSS orders are also highly spectrally concentrated.

For the waveforms we use in amplitude noise QNS, we can define a similar measure for the spectral concentration of the amplitude filter in a band $B(\omega_0)\equiv[\omega_0-\delta\omega,\omega_0+\delta\omega]$,
\begin{align*}
\mathcal{R}\equiv\frac{\int_{B(\omega_0)}d\omega\,F_\Omega(\omega,T) }{\int_{-\infty}^{\infty}d\omega\,F_\Omega(\omega,T)}. 
\end{align*}
Because the amplitude FF is defined as the \emph{continuous} Fourier transform of the amplitude waveform, the bound on the lower integral is $(-\infty,\infty)$ rather than $(-\omega_N,\omega_N)$.

\section{Details of Numerical Optimization}
\label{app:Optimization}
The dephasing-robust waveforms were originally obtained numerically, through the use of the constrained non-linear optimization solver Casadi and backend IPOPT \cite{casadi, IPOPT, gnu-parallel}.
Here we describe how we expressed the problem of filter design with bounded controls as a scalable numerical optimization problem. The procedure may be of independent interest for linear programming optimization problems with a large number of affine constraints.
The true feasible region is defined by the following set of $2 N$ linear constraints 
\begin{align}
\label{eq:appendix-a-constraints}
\pm \sum_{k=0}^{K}
v_{m}^{(k)}(N,W)
\Big[\Omega_\text{c}^{(k)}\cos[\omega_0 m \Delta t] + \Omega_\text{s}^{(k)}\sin[\omega_0 m \Delta t]\Big] \leq  \Omega_\textrm{max},\,
  \quad \forall m\in\{0,\ldots,N-1\}
 .
\end{align}

For notational convenience, we first define the $2K$ variables $x^{(k)}$ and $4KN$ coefficients $a_{m}^{(k)}$ by
\begin{align*}
    x^{(k)} &= \Omega_\text{c}^{(k)} \\
    x^{(k + K)} &= \Omega_\text{s}^{(k)} \\
    a_{m}^{(k)} &= v_{m}^{(k)}(N,W) \cos[\omega_0 m \Delta t] / \Omega_\textrm{max}  \\
    a_{m}^{(k + K)} &= v_{m}^{(k)}(N,W) \sin[\omega_0 m \Delta t] / \Omega_\textrm{max}  \\
    a_{m + N}^{(k)} &= -v_{m}^{(k)}(N,W) \cos[\omega_0 m \Delta t] / \Omega_\textrm{max}  \\
    a_{m + N}^{(k + K)} &= -v_{m}^{(k)}(N,W) \sin[\omega_0 m \Delta t] / \Omega_\textrm{max} 
    .
\end{align*}

We now recast the problem into a standard form, which can be done by dividing each inequality by $\Omega_\textrm{max}$ and combining the coefficients to produce
\begin{align}
\label{eq:appendix-general-constraints}
\sum_{k=0}^{2K}
a_{m}^{(k)} x^{(k)} \leq  1\,
  \quad \forall m \in\{0,\ldots,2N-1\}
 .
\end{align}

We are inspired by a classic routine in linear programming for identifying redundant constraints, which have the defining property that they can be pruned from the system without changing the feasible region \cite{estinmgsih2019some}.
Namely, attempt to maximize the violation of one selected constraint $m'$, subject to all the other constraints. The selected constraint is redundant with respect to the other constraints if and only if the maximum violation is negative. 

\begin{equation}
\begin{aligned}
\max_{x^{(k)}} \quad & v= - 1 + \sum_{k=0}^{2K}
a_{m'}^{(k)} x^{(k)}
\\
\textrm{s.t.} \quad &
\forall m \neq m'
\quad  \sum_{k=0}^{2K}
a_{m}^{(k)} x^{(k)} \leq  1
\end{aligned}
\label{eq:redundant}
\end{equation}

\noindent
If the result of this optimization is a positive number $v$, then there exists an assignment $x^{(k)}$ such that 
\begin{equation}
    \forall m \neq m' \quad \sum_{k=0}^{2K}
a_{m}^{(k)} x^{(k)} \leq  1,
\label{eq:appendix-constraints-without-mp}
\end{equation}
but $\sum_{k=0}^{2K}
a_{m'}^{(k)} x^{(k)} > 1$. Such an assignment lies outside the feasible region of the constraint set \eqq{eq:appendix-general-constraints}, but within the feasible region of the constraint set \eqq{eq:appendix-constraints-without-mp}. This proves that the constraint $m'$ has an effect on the feasible region if removed, and is therefore not redundant.

In contrast, if the result of the optimization is a negative number $v$, then there does not exist an assignment $x^{(k)}$ which simultaneously satisfies \eqq{eq:appendix-constraints-without-mp} while not satisfying $\sum_{k=0}^{2K}
a_{m'}^{(k)} x^{(k)} \leq 1$. By applying De Morgan's laws of formal logic, this is equivalent to the statement that \eqq{eq:appendix-constraints-without-mp} logically implies $(\sum_{k=0}^{2K}
a_{m'}^{(k)} x^{(k)} \leq 1)$.
Therefore, incorporating the constraint $m'$ does not remove any points from the feasible region. As additional constraints can never increase the size of the feasible region, this proves that the feasible regions of constraint set Eq.
\ref{eq:appendix-general-constraints}, and constraint set \eqq{eq:appendix-constraints-without-mp} are equal. Therefore, the constraint $m'$ is redundant with respect to all the other constraints and can be pruned from the system without affecting the feasible region. This can be desirable due to the gain in efficiency of having a problem with fewer constraints. For the edge-cases where the result of the optimization is zero [positive infinity], then the constraint $m'$ is redundant [not redundant], respectively.

We modify this routine to also prune \textit{approximately} redundant constraints which have a small effect on the feasible region. The algorithm first creates an overapproximation to the feasible region whose distance under the $L_{\infty}$ norm (i.e. the maximum absolute distance) from the true feasible region is bounded by $\epsilon$, and then contracts the overapproximation by a factor of $\epsilon$, thus producing an underapproximation to the true feasible region. 

For our target application, all of our constraints in our applications have a natural length-scale because they are affine constraints with a nonzero constant term. This allows us to define a ``small'' violation in terms of the constant term. Namely, a violation is small if the maximum violation of \eqq{eq:redundant} is less than a chosen $\epsilon > 0$.
This approach would be ill-defined if we included constraints with no constant term such as
$$
\sum_{k=0}^{2K}
a_{m}^{(k)} x^{(k)} \leq  0.
$$
Any violation to this constraint can be made arbitrarily large or small by simply multiplying $a_{m}^{(k)} \;\forall\; k$ by a suitable constant $>0$. This constant rescaling trick cannot be applied to constraints with a constant term, as the conversion to standard form would remove any such scaling.
As such, the question of whether a violation of this constraint is large or small depends not on the projective space of equivalent representations of this constraint, but rather on the particular way that the constraint is written. Therefore, the notion of a ``small'' violation would be ill-defined.

In addition to using this property to identify \textit{approximately} redundant constraints, we also exploit this property to later contract the feasible region by rescaling the constant term. This is used to convert an overapproximation of the feasible region into an underapproximation in an unbiased manner.
Namely, after we have identified a core subset of constraints $M$, we then tighten the constraints by a factor of $(1 + \epsilon)$ as follows

\begin{align}
\label{eq:appendix-reduced-constraints}
\sum_{k=0}^{2K}
a_{m}^{(k)} x^{(k)} \leq  (1 + \epsilon)^{-1}
  \quad \forall m \in \{0,\ldots,2N-1\}
 .
\end{align}

\begin{algorithm}[H]
\begin{algorithmic}[1]
\Require $\epsilon > 0$, affine constraints set $\mathcal C$
\State Normalize the affine constraints such that the constant term is 1
\State Randomly shuffle order of constraints
\State active set $\mathcal A \gets \emptyset$
\ForEach {$c \in \mathcal C$}
\State violation $v \gets$ maximize violation of $c$ subject to $\mathcal A$ \Comment using linear programming subroutine
\If{$v > \epsilon$}
    \State $\mathcal A \gets \mathcal A \cup c$ \Comment include non-redundant constraints
\EndIf
\EndFor
\State Tighten all constraints in $\mathcal A$ by factor of $(1 + \epsilon)$
\State Randomly shuffle order of active set $\mathcal A$
\ForEach {$c \in \mathcal A$}
\State violation $v \gets$ maximize violation of $c$ subject to $\mathcal A \setminus c$ \Comment using linear programming subroutine
\If{$v \leq 0$}
    \State $\mathcal A \gets \mathcal A \setminus c$ \Comment prune redundant constraints
\EndIf
\EndFor
\State \Return $\mathcal A$
\caption{Prune Approximatively Redundant Affine Constraints} \label{lin-prog-alg}
\end{algorithmic}
\end{algorithm}

The heuristic runtime of Algorithm 1 scales as $\mathcal{O}( m \cdot f(n, \lvert  \mathcal A\rvert ) )$, where $m$ is the number of linear inequality constraints, $\mathcal A$ is the largest active set encountered by the algorithm, and $f(n, m)$ is the time to solve a linear programming problem with $n$ variables and $m$ inequality constraints. Significantly, $\lvert \mathcal A \rvert$ does not asymptotically depend on $m$ for well-behaved instances (such as in our use-case when the constraints are continuously parameterized), making the overall runtime of Algorithm 1 linear in $m$.

For our application in Section III.B, the na\"ive formulation involved a non-linear program in 6 variables with 40000 linear inequality constraints. After the reduction with $\epsilon=0.10$, we had 6 variables with 200 linear inequality constraints. This substantially improved the efficiency of solving the optimization problem \eqq{eq:NLP}, which was otherwise intractable to run at scale.

\section{Derivation of the dephasing-robust dephasing FF}\label{app::DephasingFF}
Recall that the dephasing FF is defined by 
\begin{equation}
\label{eq:dephasing_FF}
    F_Z(\omega,T) = \left|\int_0^T\!\!\!\dd t \cos\Theta(t)e^{-i\omega t}\right|^2 + \left|\int_0^T\!\!\!\dd t \sin\Theta(t)e^{-i\omega t}\right|^2,
\end{equation}
where $\Theta(t)=\int_0^t ds\Omega(s)$ is the angle of rotation about $\sigma_1$ generated by the control.
For $\Omega(t)=\Omega_0\sin(\lambda t)$, the angle rotation becomes $\Theta(t)=\Omega_0(1-\cos\lambda t)/\lambda$. Observe that $\Theta(t)$, like $\Omega(t)$, is periodic over $\tau_\lambda\equiv 2\pi/\lambda$. If the total evolution time is an integer multiple of $\tau_\lambda$, i.e. $T=M\tau_\lambda$, we can use the periodic nature of $\Theta(t)$ to write the cosine term in \eqq{eq:dephasing_FF} as
\begin{align}\notag
\left|\int_0^T\!\!\!\dd t \cos\Theta(t)e^{-i\omega t}\right|^2
=&\,\bigg|\sum_{m=0}^{M-1}\int_{m\tau_\lambda}^{(m+1)\tau_\lambda}\!\!\!\!\!\!\!\dd t
\cos\Theta(t)e^{-i\omega t}\bigg|^2\\
=&\,\bigg|\sum_{m=0}^{M-1}e^{-i m\omega\tau_\lambda}     \bigg|^2\;\bigg|\int_0^{\tau_\lambda}\!\!\dd t\cos\Theta(t)e^{-i\omega t}\bigg|^2.\label{eq::CosineTerm}
\end{align}
Summing the geometric series in the first term produces \cite{AlvarezPRL2011}
\begin{align}
\bigg|\sum_{m=0}^{M-1}e^{-i m\omega\tau_\lambda}     \bigg|^2=\frac{\sin^2(M\pi\omega/\lambda)}{\sin^2(\pi\omega/\lambda)}.\label{eq::comb1}
\end{align}
By subsituting this expression into \eqq{eq::CosineTerm} and performing a similar calculation on the sine term in \eqq{eq:dephasing_FF}, we find
\begin{align}
&\left|\int_0^T\!\!\!\dd t \cos\Theta(t)e^{-i\omega t}\right|^2=\frac{\sin^2(M\pi\omega/\lambda)}{\sin^2(\pi\omega/\lambda)}\bigg|\int_0^{\tau_\lambda}\!\!\dd t\cos\Theta(t)e^{-i k\lambda t}\bigg|^2,\label{eq::CosineTerm2}\\
&\left|\int_0^T\!\!\!\dd t \sin\Theta(t)e^{-i\omega t}\right|^2=\frac{\sin^2(M\pi\omega/\lambda)}{\sin^2(\pi\omega/\lambda)}\bigg|\int_0^{\tau_\lambda}\!\!\dd t\sin\Theta(t)e^{-i k\lambda t}\bigg|^2. \label{eq::SineTerm2}
\end{align}
Subsituting these expression into \eqq{eq:dephasing_FF} produces the exact form of the dephasing FF given in \eqq{magical_deph_ff_1} of the main text. 

To obtain \eqq{magical-ff}, the approximate form of the dephasing FF in terms of Bessel functions, we approximate \eqq{eq::comb1} as a frequency comb in the limit of large $M$ \cite{AlvarezPRL2011}, 
\begin{align}
\bigg|\sum_{m=0}^{M-1}e^{-i m\omega\tau_\lambda}     \bigg|^2=\frac{\sin^2(M\pi\omega/\lambda)}{\sin^2(\pi\omega/\lambda)}\;\overset{M\gg 1}{\approx}\; M\lambda \sum_{k\in \mathbb{Z}}\delta(\omega-k\lambda). \label{eq::comb}
\end{align}
When $M\gg 1$, cosine and sine terms in Eqs. (\ref{eq::CosineTerm2}) and (\ref{eq::SineTerm2}) then take the approximate forms
\begin{align}
&\left|\int_0^T\!\!\!\dd t \cos\Theta(t)e^{-i\omega t}\right|^2\approx M\lambda \sum_{k\in \mathbb{Z}}\delta(\omega-k\lambda)\bigg|\int_0^{\tau_\lambda}\!\!\dd t\cos\Theta(t)e^{-i k\lambda t}\bigg|^2,\label{eq::CosineTerm3}\\
&\left|\int_0^T\!\!\!\dd t \sin\Theta(t)e^{-i\omega t}\right|^2\approx M\lambda \sum_{k\in \mathbb{Z}}\delta(\omega-k\lambda)\bigg|\int_0^{\tau_\lambda}\!\!\dd t\sin\Theta(t)e^{-i k\lambda t}\bigg|^2. \label{eq::SineTerm3}
\end{align}
 The integral terms in Eqs. (\ref{eq::CosineTerm3}) and (\ref{eq::SineTerm3}) can
 be written in terms of exponential functions,
\begin{align}\label{eq::CosineTerm4}
&\int_0^{\tau_\lambda}\!\!\dd t\cos\Theta(t)e^{-i k\lambda t}=\frac{1}{2}\int_0^{\tau_\lambda}\!\!\dd t e^{i\Theta(t)}e^{-i k\lambda t}+
\frac{1}{2}\int_0^{\tau_\lambda}\!\!\dd t e^{-i\Theta(t)}e^{-i k\lambda t},\\
&\int_0^{\tau_\lambda}\!\!\dd t\sin\Theta(t)e^{-i k\lambda t}=\frac{1}{2i}\int_0^{\tau_\lambda}\!\!\dd t e^{i\Theta(t)}e^{-i k\lambda t}-
\frac{1}{2i}\int_0^{\tau_\lambda}\!\!\dd t e^{-i\Theta(t)}e^{-i k\lambda t}.\label{eq::SineTerm4}
\end{align}
By substituting $\tau_\lambda=2\pi/\lambda$ and $\Theta(t)=\Omega_0(1-\cos\lambda t)/\lambda$ into the positive exponential term, we find
\begin{align*}
\int_0^{\tau_\lambda}\!\!\dd t e^{i\Theta(t)}e^{-i k\lambda t}=&\,
e^{i\Omega_0/\lambda}\int_0^{2\pi/\lambda}\!\!\dd t e^{-i\big(\frac{\Omega_0}{\lambda}\cos\lambda t+k\lambda t\big)}\\
=&\,
\frac{e^{i\Omega_0/\lambda}}{\lambda}\int_0^{2\pi}\!\!\dd \tau e^{-i\big(\frac{\Omega_0}{\lambda}\cos\tau+k\tau\big)}\\
=&\,
\frac{e^{i\Omega_0/\lambda}e^{ik\pi/2}}{\lambda}
\int_{\pi/2}^{5\pi/2}\!\!\dd \tau e^{-i\big(\frac{\Omega_0}{\lambda}\sin\tau+k\tau\big)}\\
=&\,
\frac{e^{i\Omega_0/\lambda}e^{ik\pi/2}}{\lambda}
\left[\int_{0}^{2\pi}\!\!\dd \tau e^{-i\big(\frac{\Omega_0}{\lambda}\sin\tau+k\tau\big)}+\int_{2\pi}^{5\pi/2}\!\!\dd \tau e^{-i\big(\frac{\Omega_0}{\lambda}\sin\tau+k\tau\big)}-\int_{0}^{\pi/2}\!\!\dd \tau e^{-i\big(\frac{\Omega_0}{\lambda}\sin\tau+k\tau\big)}\right].
\end{align*}
The last two terms in this expression cancel, leaving
\begin{align}
\int_0^{\tau_\lambda}\!\!\dd t e^{i\Theta(t)}e^{-i k\lambda t}=&\,
\frac{e^{i\Omega_0/\lambda}e^{ik\pi/2}}{\lambda}\int_{0}^{2\pi}\!\!\dd \tau\, e^{-i\big(\frac{\Omega_0}{\lambda}\sin\tau+k\tau\big)}\notag\\
=&\,\frac{e^{i\Omega_0/\lambda}e^{-ik\pi/2}}{\lambda}\int_{-\pi}^{\pi}\!\!\dd \tau\, e^{i\big(\frac{\Omega_0}{\lambda}\sin\tau-k\tau\big)}\notag
\\
=&\,\frac{2\pi e^{i\Omega_0/\lambda}e^{-ik\pi/2}}{\lambda}J_k\!\left(\frac{\Omega_0}{\lambda}\right),\label{eq::Term1}
\end{align}
where the last equality follows from the integral representation of a Bessel function of the first kind.
Performing a similar calculation on the negative exponential term yields
\begin{align}
\int_0^{\tau_\lambda}\!\!\dd t e^{-i\Theta(t)}e^{-i k\lambda t}=\frac{2\pi e^{-i\Omega_0/\lambda}e^{-ik\pi/2}}{\lambda}J_{-k}\!\left(\frac{\Omega_0}{\lambda}\right)
=\frac{2\pi e^{-i\Omega_0/\lambda}e^{-ik\pi/2}}{\lambda}(-1)^k J_{k}\!\left(\frac{\Omega_0}{\lambda}\right).\label{eq::Term2}
\end{align}

By combining Eqs. (\ref{eq:dephasing_FF}), (\ref{eq::CosineTerm3}), (\ref{eq::SineTerm3}),
(\ref{eq::CosineTerm4}), (\ref{eq::SineTerm4}), (\ref{eq::Term1}) and (\ref{eq::Term2}), the dephasing FF becomes 
\begin{align*}
    F_Z(\omega,T)\approx&\,M\lambda\sum_{k\in \mathbb{Z}}\delta(\omega-k\lambda)\left[\;\bigg|\int_0^{\tau_\lambda}\!\!\dd t\cos\Theta(t)e^{-i k\lambda t}\bigg|^2+\bigg|\int_0^{\tau_\lambda}\!\!\dd t\sin\Theta(t)e^{-i k\lambda t}\bigg|^2\;\right]\\
    =&\,M\lambda \sum_{k\in \mathbb{Z}}\delta(\omega-k\lambda)\left[\;\frac{1}{2}\bigg|\int_0^{\tau_\lambda}\!\!\dd t e^{i\Theta(t)} e^{-i k\lambda t}\bigg|^2+\frac{1}{2}\bigg|\int_0^{\tau_\lambda}\!\!\dd t e^{-i\Theta(t)} e^{-i k\lambda t}\bigg|^2\;\right]\\
    =&\,M\lambda\left(\frac{2\pi}{\lambda}\right)^2 \sum_{k\in \mathbb{Z}}\delta(\omega-k\lambda)\;\bigg|J_k\!\left(\frac{\Omega_0}{\lambda}\right)\bigg|^2\\
    =&\,2\pi T \sum_{k\in \mathbb{Z}}\delta(\omega-k\lambda)\;\bigg|J_k\!\left(\frac{\Omega_0}{\lambda}\right)\bigg|^2
\end{align*}
Note that the same approximate dephasing FF can be obtained for the waveform $\Omega(t)=\Omega_0\cos(\lambda t)$.

\section{Higher Order FF Computation} \label{app::DFT}

In Appendix \ref{app::Measurements}, an equation for the higher-order dephasing filter function $G_Z(\omega,\omega',T)$ was derived.

\begin{align}
&G_Z(\omega,\omega',T)=\int_0^{T}\!\!\!\!dt_1\!\!\int_0^{t_1}\!\!\!\!dt_2\,\sin\big[\Theta(t_1)\!-\!\Theta(t_2)\big] \label{eq:B1GZ} \\
&\;\;\;\times\int_0^{T}\!\!\!\!dt_3\!\!\int_0^{t_3}\!\!\!\!dt_4\,\sin\big[\Theta(t_3)\!-\!\Theta(t_4)\big]\notag\\&\;\;\;\big[e^{i\omega(t_1-t_2)}e^{i\omega'(t_3-t_4)}+e^{i\omega(t_1-t_3)}e^{i\omega'(t_2-t_4)}
+e^{i\omega(t_1-t_4)}e^{i\omega'(t_2-t_3)}\big].\notag
\end{align}

This formula involves a quadruply-nested integral, and it is not obvious how to factor the expression into smaller, more manageable integrals. Directly using this expression to evaluate the higher-order filter function at single pair of frequencies $\omega, \omega'$ has a runtime complexity of $\mathcal{O}(N^4)$, where $N$ is the number of segments in the piecewise-constant waveform. To sample the higher-order FF at a grid of frequencies (with density proportional to $N$) has complexity $\mathcal{O}(N^6)$, which is extremely computationally intensive.
Here we describe a more efficient approach for sampling $G_Z(\omega,\omega',T)$ at a grid of frequencies with runtime $\mathcal{O}(N^2 \log(N)^2)$.

First, we use the sum and difference trigonometric identities to rewrite $\sin\big[\Theta(t_1)\!-\!\Theta(t_2)\big]$ into a (sum of) products
\begin{equation}
    \sin\big[\Theta(t_1)\!-\!\Theta(t_2)\big] = \sin \Theta(t_1) \cos \Theta(t_2) - \cos \Theta(t_1) \sin \Theta(t_2) .
\end{equation}

Using this to rewrite \eqq{eq:B1GZ}, we get

\begin{align*}
&G_Z(\omega,\omega',T)=
\int_0^{T}dt_1\int_0^{t_1}dt_2
\int_0^{T}dt_3\int_0^{t_3}dt_4
\\
&\big[
\sin \Theta(t_1) \cos\Theta(t_2) - \cos \Theta(t_1) \sin\Theta(t_2) \big] \\ 
\times
&\big[ \sin \Theta(t_3) \cos\Theta(t_4) - \cos \Theta(t_3) \sin\Theta(t_4) \big]
\\
\times &\big[
e^{i\omega(t_1-t_2)}e^{i\omega'(t_3-t_4)}
 + e^{i\omega(t_1-t_3)}e^{i\omega'(t_2-t_4)}
 + e^{i\omega(t_1-t_4)}e^{i\omega'(t_2-t_3)}
\big].
\end{align*}

This expression can be expanded into a sum of 12 integrands.  For the remainder of this section, we shall focus on the following subexpression, which has the arguments in the exponentials split according to the groupings $t_1,t_4$ and $t_2, t_3$ (which we note is different than the groupings $t_1,t_2$ and $t_3, t_4$ induced by the bounds on integration). The other subexpressions proceed similarly.

\begin{equation}
f(\omega, \omega') =
\int_0^{T}dt_1\int_0^{t_1}dt_2
\int_0^{T}dt_3\int_0^{t_3}dt_4
\sin \Theta(t_1) \cos\Theta(t_2)
\sin \Theta(t_3) \cos\Theta(t_4)
e^{i\omega(t_1-t_4)}e^{i\omega'(t_2-t_3)}
\label{eq:hard-subexpr}
\end{equation}

For convenience, we define a helper function $H(\omega_1, \omega_2)$, where we emphasize the notational change from $(\omega,\omega')$ to $(\omega_1, \omega_2)$

\begin{equation}
H(\omega_1, \omega_2) = 
\int_0^{T}dt_1\int_0^{t_1}dt_2
\sin \Theta(t_1) \cos\Theta(t_2)
e^{i\omega_1 t_1}e^{i\omega_2 t_2}.
\end{equation}

We can now begin to rewrite \eqq{eq:hard-subexpr}
\begin{align*}
f(\omega, \omega') &=
\int_0^{T}dt_1\int_0^{t_1}dt_2
\int_0^{T}dt_3\int_0^{t_3}dt_4
\sin \Theta(t_1) \cos\Theta(t_2)
\sin \Theta(t_3) \cos\Theta(t_4)
e^{i\omega(t_1-t_4)}e^{i\omega'(t_2-t_3)}
\\
&=
\int_0^{T}dt_1\int_0^{t_1}dt_2
\int_0^{T}dt_3\int_0^{t_3}dt_4
\sin \Theta(t_1) \cos\Theta(t_2)
\sin \Theta(t_3) \cos\Theta(t_4)
e^{i\omega t_1} e^{-i\omega t_4} e^{i\omega't_2} e^{-i\omega't_3}
\\
&=
\left(
\int_0^{T}dt_1\int_0^{t_1}dt_2
\sin \Theta(t_1) \cos\Theta(t_2)
e^{i\omega t_1} e^{i\omega' t_2}
\right)
\left(
\int_0^{T}dt_3\int_0^{t_3}dt_4
\sin \Theta(t_3) \cos\Theta(t_4)
e^{-i\omega' t_3} e^{-i\omega t_4}
\right)
\\
&=
H(\omega, \omega') \cdot H(-\omega', -\omega).
\end{align*}

As such, we see that the arguments in the exponentials depending on different time orderings is no great barrier to factoring the subexpression into doubly-nested integrals. Indeed, all 12 subexpressions in our expanded formula for $G_Z(\omega,\omega',T)$ yield to such decompositions (some using different helper functions). Importantly, decomposing the subexpression in this manner reduces the complexity from evaluating a single quadruply-nested integral to computing two doubly-nested integrals, which are substantially faster to calculate (runtime $\mathcal{O}(N^2)$ as opposed to $\mathcal{O}(N^4)$). Through this approach, we can compute a single element $G_Z(\omega,\omega',T)$ in runtime $\mathcal{O}(N^2)$ by computing many doubly-nested integrals.

\subsection{Discrete Fourier Transform to Efficiently Compute Higher Order FF}
We have shown how to compute a single element $G_Z(\omega,\omega',T)$ in runtime $\mathcal{O}(N^2)$ by computing many doubly-nested integrals. Next, we demonstrate how to use the fast-fourier transform to compute $G_Z(\omega,\omega',T)$ for many frequencies at once, with total runtime $\mathcal{O}(N^2 \log(N)^2)$.

We focus on the computationally-relevant Discrete Fourier Transform (DFT) application, where we discretize the function $\Theta(t)$ into a sequence of length $N$, and the frequencies $\omega, \omega'$ at which we evaluate $G_Z(\omega,\omega',T)$ are integer multiples of $\Delta \omega = 2\pi / (T N)$. It suffices to demonstrate that $H(\omega_1, \omega_2)$ can be efficiently computed at all such frequencies simultaneously (in runtime $\mathcal{O}(N^2)$). Once this has been done, then computing $G_Z(\omega,\omega',T)$ is straightforward.

We begin be rewriting

\begin{align*}
H(\omega_1, \omega_2) &= 
\int_0^{T}dt_1\int_0^{t_1}dt_2
\sin \Theta(t_1) \cos\Theta(t_2)
e^{i\omega_1 t_1}e^{i\omega_2 t_2}
\\
&= \int_0^{T}dt_1 e^{i\omega_1 t_1} \sin \Theta(t_1) 
\int_0^{t_1}dt_2 \cos\Theta(t_2) e^{i\omega_2 t_2}
\end{align*}

Next, we define a few more helper functions
\begin{align*}
    h(t, \omega) &\equiv \int_0^{t}ds \cos\Theta(s) e^{i\omega s}
\\
&= \int_0^{T}ds\ g(t, s) e^{i\omega s}
,
\\
\textrm{where } g(t, s) &\equiv 
     \begin{cases} 
      \cos\Theta(s) & s \leq t \\
      0 & s > t.
   \end{cases}
\end{align*}

Then we discretize time and convert the integrals to finite sums

\begin{align}
H(\omega_1, \omega_2) &\equiv (\Delta t)^2
\sum_{j_1=0}^{N-1} 
e^{i\omega_1 j_1 \Delta t} \sin \Theta(j_1 \Delta t) 
\sum_{i_2=0}^{j_1} \cos\Theta(j_2 \Delta t) e^{i\omega_2 j_2 \Delta t},
\label{eq:helper-dft}
\\
h(k, \omega) &\equiv (\Delta t) \sum_{j=0}^{N-1} \ g(k, j) e^{i\omega j \Delta t},
\label{eq:g-dft}
\\
\textrm{where } g(k, j) &\equiv
    \begin{cases} 
    \cos\Theta(j) & j \leq k \\
    0 & j > k.
    \end{cases}
\label{eq:g-def}
\end{align}

If $k$ is regarded as a fixed parameter, then \eqq{eq:g-dft} is readily seen to be the discrete fourier transform of $g(k, j)$ in \eqq{eq:g-def} with respect to $j$.
Therefore for each fixed value of $t$, the DFT is able to compute $h(k, \omega)$ for all (discretized) values of $\omega = m \cdot \Delta \omega$ simultaneously in time $\mathcal{O}(N \log(N))$. By repeating this for each (discretized) time $t$, we are able to compute all values of $h(k, \omega)$ in runtime $\mathcal{O}(N^2 \log(N)^2)$.

After we have pre-computed all values of $h(t, \omega)$, we express the original helper function $H(\omega_1, \omega_2)$ as

\begin{align*}
H(\omega_1, \omega_2) &\equiv (\Delta t)^2
\sum_{j_1=0}^{N-1} 
e^{i\omega_1 j_1 \Delta t} \sin \Theta(j_1 \Delta t) 
\sum_{i_2=0}^{j_1} \cos\Theta(j_2 \Delta t) e^{i\omega_2 j_2 \Delta t},
\\
&= (\Delta t) \sum_{j_1=0}^{N-1} 
e^{i\omega_1 j_1 \Delta t} 
\big[ \sin \Theta(j_1 \Delta t) \ h(j_1, \omega_2) \big]. \numberthis \label{eq:helper-sin-h}
\end{align*}

And we see that it can again be regarded as a Fourier transform. Explicitly, we now regard $\omega_2$ as a fixed parameter, and take the discrete Fourier transform with respect to $t_1$. This can be computed for all frequencies $\omega_1$ in runtime $\mathcal{O}(N \log(N))$ using the DFT.
By repeating this for all values of $\omega_2$ (without needing to recompute $h(k, \omega)$ from the previous step), we can thus compute $H(\omega_1, \omega_2)$ for all (discretized) frequencies $\omega_1, \omega_2$ in runtime $\mathcal{O}(N^2 \log(N)^2 )$. We note that this is almost the same asymptotic complexity as it takes to compute a single element of $H$.
Thus, by carefully rearranging our original expression we can leverage the DFT to efficiently compute the higher-order FFs.

\end{appendix}
\end{widetext}

\bibliographystyle{apsrev4-1}
\bibliography{bibliography}

\end{document}